\documentclass[pra,superscriptaddress,twocolumn,english,longbibliography,showpacs=false]{revtex4-2}

\usepackage[T1]{fontenc}
\usepackage[utf8]{inputenc}
\usepackage{xcolor}
\usepackage{babel}
\usepackage{amsmath}
\usepackage{amssymb}
\usepackage{graphicx}
\usepackage{physics} 
\usepackage{dsfont}
\usepackage{hyperref}
\usepackage{amsthm}
\usepackage{mathtools}
\usepackage[left=23mm,right=13mm,top=35mm,columnsep=15pt]{geometry} 
\usepackage{adjustbox}
\usepackage{placeins}
\usepackage{csquotes}
\usepackage{mathrsfs}
\usepackage{enumerate}
\usepackage{verbatim}
\usepackage{booktabs}
\usepackage{subcaption}
\usepackage{ragged2e}

\begin{document}

\title{Gravitational decoherence of a composite particle: the interplay between gravitons and a classical Newtonian potential}

\author{Thiago H. Moreira\href{https://orcid.org/0000-0001-7093-0287}{\includegraphics[scale=0.05]{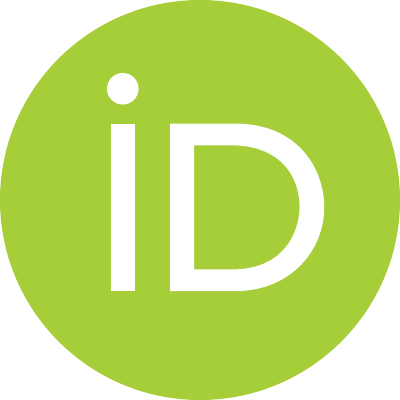}}}
\email{thiagohenriquemoreira@discente.ufg.br}
\affiliation{QPequi Group, Institute of Physics, Federal University of Goi\'as, Goi\^ania, Goi\'as, 74.690-900, Brazil}

\author{Lucas C. Céleri\href{https://orcid.org/0000-0001-5120-8176}{\includegraphics[scale=0.05]{orcidid.pdf}}}
\email{lucas@qpequi.com}
\affiliation{QPequi Group, Institute of Physics, Federal University of Goi\'as, Goi\^ania, Goi\'as, 74.690-900, Brazil}

\begin{abstract}
The fact that gravitational environments cannot be shielded (since gravity is universal) makes them of great theoretical interest to decoherence mechanisms and to the quantum-to-classical transition. While past results seemed to indicate that graviton-induced decoherence of spatial superpositions happens only for macroscopic systems, recently it was shown that this mechanism can be enhanced through the system's own dynamical internal structure. In this work, we extend this analysis by including the interaction with a classical Newtonian potential. We show that, although the graviton bath alone dominates the mechanism for short times compared to a timescale established by the size of the quantum spatial superposition, the interaction between the gravitons and the internal degrees of freedom of the system renders decoherence inevitable in the long-time limit, even for microscopic masses. We also show that this mechanism is slightly slowed down by the interplay with the classical Newtonian potential, which, for systems without dynamical internal degrees of freedom, can even lead to recoherence-like behaviour, at least in principle.

\vspace{0.5cm}
\end{abstract}


\maketitle


\section{Introduction} \label{Sec:Introduction}

Among the many known sources of decoherence in quantum systems, those related to gravitational interaction have attracted some attention due to the universal aspect of the latter. The term gravitational decoherence refers to any loss of coherence in quantum systems that is somewhat related to gravity, either directly or indirectly~\cite{Bassi_2017,Hsiang_2024,Pfister2016}. It can refer to gravity-induced collapse models~\cite{Karolyhazy1966,Diosi1984,Diosi1989,Frenkel1990,Penrose1996,Diosi2007,Diosi2014,Adler2014,Samuel2018}, self-decoherence models at a Heisenberg cut establishing how classical and quantum mechanics emerge from a yet unknown theory at the Planck scale~\cite{Aguiar2025}, decoherence due to the coupling with classical~\cite{Linet1976,Stodolsky1979,Cai1989,Power2000,Reynaud2004,Lamine2006,Breuer2009} and quantum~\cite{Blencowe_2013,Anastopoulos_2013,Kanno2021,Lagouvardos2021,HABA2002,Oniga2016,Fahn2025,Cho2025} gravitational radiation described in the linearised limit of general relativity, and even decoherence of a composite particle induced by the coupling between external and internal degrees of freedom (DoFs) due to classical gravitational time dilation~\cite{Pikovski2015,Pikovski2017,Carlesso2016}. Since the gravitational force is much weaker than the other fundamental interactions, the experimental realisation of such gravitational effects faces the problem of controlling competing sources of decoherence. Nevertheless, the interest in this topic lies in the fact that, however weak, gravitational environments cannot be shielded. In this work, we will consider decoherence induced by the interaction with quantum gravitational radiation degrees of freedom.

Despite considerable theoretical efforts~\cite{Eppley1977,Kiefer2012,Penrose2014,Salvio_2018,Percacci2017,Loll_2019,Rovelli2007,Polchinski2005,Polchinski2005_2}, the quantum nature of the gravitational interaction still lacks experimental corroboration due to its weakness relative to the other known fundamental interactions, which are satisfactorily described by quantum field theory. Regardless of which theory of quantum gravity (if any) ever proves to be the most adequate, we can expect the usual quantum field theoretical treatment to be suitable in the limit of weak gravitational fields. In this limit, the spacetime metric is described by an expansion of the gravitational propagating DoFs around some known classical solution to Einstein's equation. This treatment describes the so-called classical gravitational waves, which were detected by LIGO in 2015~\cite{Abbott_2016}. At the quantum level, these waves give rise to a spin 2 excitation called the graviton. This formalism, referred to as perturbative quantum gravity, faces problems concerning perturbative non-renormalizability~\cite{Basile2025}. Nevertheless, it stands as a predictive effective field theory. For example, radiative corrections to the Newtonian potential were explicitly computed in~\cite{Donoghue1994}, where it was found that, for a potential given by $\phi(r)=-GM/r$, these corrections are of the order of $\hbar G/c^3r^2$, which are too small to be measured with current and, probably, near-future technology.

Motivated by such practical limitations in direct detection of gravitons (if such detection is even possible in principle~\cite{Dyson_2013,Carney2024,Tobar2024}), different ways of detecting quantum aspects of gravity in more indirect ways have been proposed in the last years~\cite{Carney2019,Bose2025,Beyer2025}. Examples include gravity-induced entanglement~\cite{Bose2017,Marletto2017,Carney2021,Danielson_2022,Christodoulou2023,Christodoulou2023b}, graviton noise affecting the geodesic deviation of test particles~\cite{Parikh2020,Parikh_2021,Parikh2021,Cho2022,Cho_2023,Chawla2023}, and, of course, graviton-induced decoherence.

In particular, the decoherence of spatial superpositions of a quantum point particle induced by the interaction with a graviton bath was analysed in~\cite{Kanno2021}, where the authors were led to the conclusion that the off-diagonal elements of the reduced density matrix of the system only go through significant decay for systems with momentum greater than the Planck mass, $M_{\rm P}\sim10^{-8}$ kg. Now, the most massive quantum systems to ever have been put into spatial superposition are molecules with $m\sim10^{-22}$ kg~\cite{Gerlich2011,Fein2019,Pedalino2025}, which illustrates the immense challenge of observing such effects. However, as pointed out in~\cite{Kanno2021}, the decoherence rate can be enhanced by considering different configurations of the superposition state and also other possible initial graviton states.

In Ref.~\cite{Moreira_2023}, this decoherence mechanism was analysed in the context of a quantum composite particle, namely a system described by both external and internal DoFs. Due to the universal aspect of gravity, the gravitons couple with all dynamical DoFs of the system. For quantum superpositions of the external variables, the internal DoFs enter as a second environment, which interacts with the gravitational one. The interplay between the system internal structure and the gravitons (which does not amount to a simple sum of their individual effects due to their mutual interaction) was found to significantly enhance decoherence, allowing for the possibility of graviton-induced spatial localisation of quantum superpositions involving microscopic masses. An interesting application of the results for graviton-induced decoherence was analysed in~\cite{Moreira2024}, where the entropy production due to the interaction with the graviton bath was quantified by the establishment of a fluctuation theorem (see~\cite{Basso2023,Basso2025,Costa2025} for the classical gravitational field results).

In this work, we extend the analysis of Ref.~\cite{Moreira_2023} to include the interaction of the composite system with a classical Newtonian gravitational potential, in addition to the graviton bath. While both gravitational interactions induce a coupling between the external and internal DoFs of the system, the graviton bath also acts as a quantum environment, and we shall see how the classical gravitational potential affects the graviton-induced decoherence mechanism. 

Just as in Ref. \cite{Moreira_2023}, the analysis is performed by considering four different initial states of the graviton bath. The most natural choice is the vacuum state, but we also consider an initial thermal state, for example. Throughout this work, we refer to a parameter $T_{\rm g}$ as the graviton temperature, which characterises the thermal initial state. However, we emphasize that the parameter $T_g$ should not necessarily be interpreted as the thermodynamic temperature of an equilibrated graviton gas. In the present context, it is more appropriately understood as an effective noise-temperature parameter that fixes the occupation numbers of the initial Gaussian thermal state of the graviton modes and, consequently, the power spectral density entering the gravitational noise kernel. This distinction is relevant because gravitons interact extremely weakly and cannot generally be assumed to thermalize on physically reasonable timescales. Therefore, the thermal state considered here should be regarded as a phenomenological model for a stationary gravitational noise background with a Planck-like spectrum, rather than as a claim that the gravitational radiation field is necessarily in genuine thermodynamic equilibrium~\cite{Anastopoulos_2013}. Additionally, we explore the analogy with quantum optics and consider gravitons in initial coherent and squeezed states. Although a coherent graviton state is expected to be the quantum-mechanical state whose properties most closely resemble those of a classical gravitational wave, it has been argued that gravitons created from quantum fluctuations in the course of cosmological evolution should now be in strongly squeezed states~\cite{Grishchuk1990}.

This work is organised as follows. In Sec.~\ref{Sec:Classical-action} we obtain the classical action describing the interaction between the system, the gravitational radiation background and the classical Newtonian potential. Since some attention is required when considering gravitational waves propagating in curved spacetime, the gravitational sector of the total action is explored in detail in the Appendix~\ref{A:Gravitons-Newton}. In Sec.~\ref{Sec:Time-evolution} we quantise both the system and the metric perturbation field, integrating over the final states of the gravitons as well as the internal DoFs of the composite particle. This is accomplished by using the Feynman-Vernon influence functional approach to open quantum systems~\cite{Feynman1963,Feynman2010,Calzetta2008}, which leads to the time evolution of the reduced density matrix describing the external DoFs of the system. The influence of the graviton bath enters the time evolution as noise, and we explicitly compute the noise kernel in the Appendix~\ref{A:Gravitational-Noise-Kernel}. Our main results are described in Sec.~\ref{Sec:Decoherence-function}, where we explicitly compute the decoherence function for a spatial superposition of the system external coordinates and compute the decoherence time for the four initial graviton states mentioned above. We also address the possibility of gravitational recoherence, which is made possible by the interplay between the graviton bath and the classical Newtonian potential. Since the explicit expressions for the decoherence functions (for the different initial states) are quite cumbersome, we list them in the Appendix~\ref{App:Explicit-expressions-dec-func}. Finally, Sec.~\ref{Sec:Conclusions} closes with some concluding remarks. Unless explicitly stated otherwise, we work with Planck units, for which $\hbar=c=G=k_B=1$. We also use the mostly-plus metric convention $\eta_{\mu\nu}=\textrm{diag}\qty(-1,+1,+1,+1)$.

\section{The classical action} \label{Sec:Classical-action}

\subsection{The matter action} \label{Sec:The-matter-action}

Let us begin by describing the system that interacts with the gravitational fields. Consider the classical action of two masses $M$ and $m$, described by variables $\zeta^\mu$ and $\xi^\mu$, respectively, in a given coordinate system, and suppose that the dynamics of the particle with mass $M$ is described only by external degrees of freedom, like the centre-of-mass coordinates, while the particle with mass $m$ is described by both internal and external DoFs. The matter action reads~\cite{Zych2019}
\begin{align} \label{Matter-action}
    S_{\rm matter}&=-M\int\dd t\,\sqrt{-g_{\mu\nu}(\zeta)\dot{\zeta}^\mu\dot{\zeta}^\nu} \nonumber \\
    &+\int\dd t\,L_{\rm rest}(\varrho,\dot{\varrho}\,\bar{t})\sqrt{-g_{\mu\nu}(\xi)\dot{\xi}^\mu\dot{\xi}^\nu},
\end{align}
where $g_{\mu\nu}$ is the spacetime metric field and
\begin{equation} \label{Rest-lagrangian}
    L_{\rm rest}(\varrho,\dot{\varrho}\,\bar{t})=-m+\mathscr{L}(\varrho,\dot{\varrho}\,\bar{t}),
\end{equation}
with $\mathscr{L}(\varrho,\dot{\varrho}\,\bar{t})$ describing the internal dynamics with coordinate $\varrho$ and generalised velocity $\dot{\varrho}=\dv*{\varrho}{t}$. We have also defined $\bar{t}=\dv*{t}{\tau}$, with $\tau$ being the proper time of mass $m$.

Let us consider the mass $M$ to be on shell with worldline $\zeta^\mu(t)=t\,\delta_0^\mu$, such that it remains at rest at the origin of our coordinate system, and with the coordinate time $t$ coinciding with its proper time. We also consider $M\gg m$. Under these assumptions, the first term in the action~\eqref{Matter-action} essentially has no dynamics, and, from now on, our system of interest is understood to be the composite particle.

\begin{figure}[!ht]
    \centering
    \includegraphics[width=0.45\linewidth]{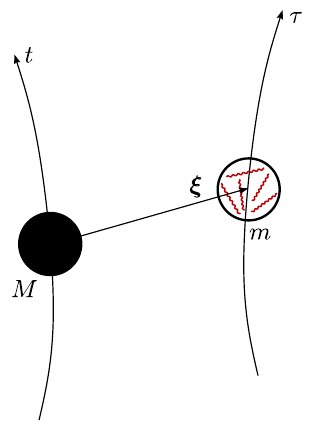}
    \caption{\justifying{Two test masses $M$ and $m$, with $M\gg m$, and their geodesic deviation in Fermi normal coordinates, represented by the vector $\boldsymbol{\xi}$. The mass $m$ is also described by internal degrees of freedom, represented by the curly red lines.}}
    \label{Fig:geodesic-deviation}
\end{figure}

Within such context, we can think of $(t,\xi^i)$ as the Fermi normal coordinates defined with respect to the worldline of the mass $M$. In these coordinates, the metric components can be written as~\cite{Manasse1963}
\begin{subequations}
\begin{align}
    g_{00}(t,\xi^i)&=-1-R_{i0j0}(t,0)\xi^i\xi^j+O\qty(\xi^3/R_0^3), \\
    g_{0i}(t,\xi^i)&=-\frac{2}{3}R_{0jik}(t,0)\xi^j\xi^k+O\qty(\xi^3/R_0^3), \\
    g_{ij}(t,\xi^i)&=\delta_{ij}-\frac{1}{3}R_{ikjl}(t,0)\xi^k\xi^l+O\qty(\xi^3/R_0^3),
\end{align}
\end{subequations}
where $R_{\mu\nu\rho\sigma}$ is the Riemann curvature tensor and $R_0$ measures the scale on which the metric changes appreciably. Physically, the use of Fermi normal coordinates allows us to interpret the coordinates $\xi^i$ as not simply describing a single particle in an arbitrary coordinate system, for which no gravitational effect could be probed due to the equivalence principle, but rather as the geodesic deviation between two test masses, one of which is massive enough to allow us to neglect its dynamics (see Figure~\ref{Fig:geodesic-deviation} for an illustration). In our parameterisation, $\xi^0(t)=t$, thus resulting
\begin{equation} \label{dot-tau}
    \sqrt{-g_{\mu\nu}(\xi)\dot{\xi}^\mu\dot{\xi}^\nu}\simeq1-\frac{1}{2}\delta_{ij}\dot{\xi}^i\dot{\xi}^j+\frac{1}{2}R_{i0j0}(t,0)\xi^i\xi^j.
\end{equation}

Next, let us specify our metric field to describe a small perturbation $h_{\mu\nu}$, which describes the gravitational radiation on some background spacetime $\Tilde{g}_{\mu\nu}$,
\begin{equation}
    g_{\mu\nu}=\Tilde{g}_{\mu\nu}+h_{\mu\nu},\hspace{0.5cm}\abs{h_{\mu\nu}}\ll\abs{\Tilde{g}_{\mu\nu}}.
\end{equation}
The Riemann tensor associated with $g_{\mu\nu}$ is given by~\cite{Wald1984}
\begin{equation}
    {R^\rho}_{\sigma\mu\nu}={\Tilde{R}^\rho}\,_{\sigma\mu\nu}+2\Tilde{\nabla}_{[\mu}{C^\rho}_{\nu]\sigma}+2{C^\rho}_{\delta[\mu}{C^\delta}_{\nu]\sigma},
\end{equation}
where $\Tilde{\nabla}_\mu$ is the covariant derivative compatible with $\Tilde{g}_{\mu\nu}$ and whose commutator defines ${\Tilde{R}^\rho}\,_{\sigma\mu\nu}$. The tensor ${C^\rho}_{\mu\nu}$ is given by
\begin{align}
    {C^\rho}_{\mu\nu}&=\frac{1}{2}g^{\rho\sigma}\qty(\Tilde{\nabla}_\mu g_{\nu\sigma}+\Tilde{\nabla}_\nu g_{\sigma\mu}-\Tilde{\nabla}_\sigma g_{\mu\nu}) \nonumber \\
    &=\frac{1}{2}\Tilde{g}^{\rho\sigma}\qty(\Tilde{\nabla}_\mu h_{\nu\sigma}+\Tilde{\nabla}_\nu h_{\sigma\mu}-\Tilde{\nabla}_\sigma h_{\mu\nu})+O(h^2),
\end{align}
which follows from metric compatibility. Then, an explicit calculation yields
\begin{eqnarray}
    R_{\rho\sigma\mu\nu}&=&\Tilde{R}_{\rho\sigma\mu\nu}+\frac{1}{2}\qty({\Tilde{R}^\lambda}\,_{\sigma\mu\nu}h_{\lambda\rho}-{\Tilde{R}^\lambda}\,_{\rho\mu\nu}h_{\sigma\lambda}) \nonumber \\
    &+&\frac{1}{2}\left(\Tilde{\nabla}_\mu\Tilde{\nabla}_\sigma h_{\rho\nu}-\Tilde{\nabla}_\mu\Tilde{\nabla}_\rho h_{\nu\sigma}\right.\nonumber\\
    &-&\left.\Tilde{\nabla}_\nu\Tilde{\nabla}_\sigma h_{\rho\mu}+\Tilde{\nabla}_\nu\Tilde{\nabla}_\rho h_{\mu\sigma}\right),
\end{eqnarray}
up to first order in $h_{\mu\nu}$.

Next, let us specialise to the case where the background itself can be seen as a small perturbation of Minkowski spacetime
\begin{equation}
    \Tilde{g}_{\mu\nu}=\eta_{\mu\nu}+h^{(B)}_{\mu\nu}.
\end{equation}
In addition, assume that this small perturbation is static and satisfies $h^{(B)}_{0i}=0$. Under these conditions, we can choose the metric perturbation $h_{\mu\nu}$ to be in the transverse-traceless (TT) gauge, Eqs.~\eqref{TT-gauge}, as shown in Appendix~\ref{A:Gravitons-Newton}. Ultimately, we are interested in considering the background metric in the Newtonian limit, for which $h^{(B)}_{\mu\nu}=-2\phi\delta_{\mu\nu}$, with $\phi(\xi)$ being the gravitational potential. In that case, we find
\begin{equation}
    \Tilde{R}_{i0j0}(\xi)=\partial_j\partial_i\phi(\xi).
\end{equation}
Putting everything together in the interaction term yields
\begin{equation} \label{Rxixi-2}
    R_{i0j0}(t,0)\xi^i\xi^j=-\qty(T_{ij}+\frac{1}{2}{T^k}_ih_{kj}+\eval{\frac{1}{2}\Ddot{h}_{ij}}_{\xi=0})\xi^i\xi^j,
\end{equation}
where $T_{ij}=-\eval{\qty(\partial_i\partial_j\phi)}_{\xi=0}$ is the tidal tensor~\cite{Hartle2003,Cho_2023}.

\begin{figure}[!ht]
    \centering
    \includegraphics[width=0.45\linewidth]{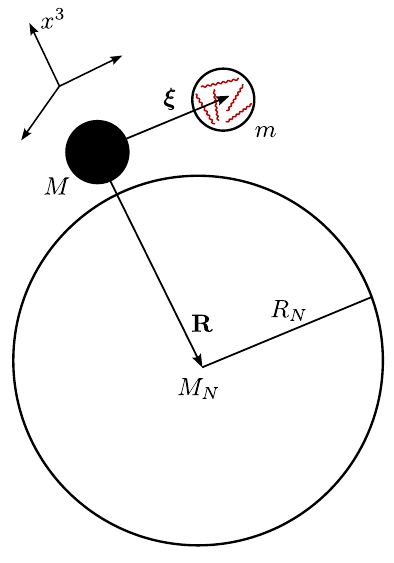}
    \caption{\justifying{The two test masses $M$ and $m$ are in the vicinity of a much bigger and much more massive spherical mass $M_N$, with radius $R_N$, located at $\vb{R}\simeq R_N\vu{e}_3$ with respect to the mass $M$.}}
    \label{Fig:Newtonian-source}
\end{figure}

In this work, we consider the two test masses to be close to a spherically symmetric Newtonian source, like the Earth, such that the much lighter one is under the influence of the gravitational potential given by
\begin{equation} \label{Newtonian-potential}
    \phi(\boldsymbol{\xi})=-\frac{M_N}{\abs{\boldsymbol{\xi}-\vb{R}}},
\end{equation}
with $M_N$ being the mass of the source and $\vb{R}$ being the radius vector that points from the mass $M$ (the origin of our coordinate system) to its centre, such that $\abs{\vb{R}}\simeq R_N$, which is its radius. Then we find the tidal tensor to take the form
\begin{equation} \label{Tidal-tensor}
    T_{ij}=\frac{M_N}{R_N^3}\qty(3\delta_{i3}\delta_{j3}-\delta_{ij}),
\end{equation}
where we chose coordinates such that $R_i\simeq R_N\delta_{i3}$ (see Figure \ref{Fig:Newtonian-source}).

Note from Eq.~\eqref{Rxixi-2} that the metric expansion in Fermi normal coordinates holds as long as $\xi^2\ll R_N^3/M_N$, and also $\xi^2\ll\omega^{-2}$, with $\omega$ denoting the angular frequency of the incoming waves. This means that we need to introduce a physical energy cutoff for incident gravitational radiation, $\Lambda\sim L_0^{-1}$, where $L_0$ is some typical geodesic separation, also referred to as the "detector size". This necessity will become evident when we quantise the gravitational radiation degrees of freedom and compute the noise kernels.

Finally, using Eqs.~\eqref{dot-tau} and~\eqref{Rxixi-2}, the action of matter becomes
\begin{align} \label{Matter-action-2}
    S_{\rm matter}&=\int\dd t\,\left\{ \frac{1}{2}m\delta_{ij}\dot{\xi}^i\dot{\xi}^j+\mathscr{L}(\varrho,\dot{\varrho}\Bar{t})\right. \nonumber \\
    &-\frac{1}{2}\mathscr{L}(\varrho,\dot{\varrho}\Bar{t})\delta_{ij}\dot{\xi}^i\dot{\xi}^j \nonumber \\
    &-\frac{1}{2}L_{\rm rest}(\varrho,\dot{\varrho}\Bar{t})T_{ij}\xi^i\xi^j \nonumber \\
    &-\frac{1}{4}h_{ij}(t,0)L_{\rm rest}(\varrho,\dot{\varrho}\Bar{t}){T^i}_k\xi^k\xi^j \nonumber \\
    &\left. -\frac{1}{4}h_{ij}(t,0)\dv[2]{t}\qty[L_{\rm rest}(\varrho,\dot{\varrho}\Bar{t})\xi^i\xi^j]\right\} .
\end{align}
The first line describes the free terms, while the second one describes the interaction between the internal and external DoFs induced by the flat spacetime metric. Their interaction with the classical Newtonian potential is on the third line, while the fourth and fifth lines describe their interaction with the metric perturbation field.

\subsection{The graviton action}

We show in Appendix~\ref{A:Gravitons-Newton} that the interaction between gravitons and the Newtonian potential is described by the action
\begin{equation} \label{Graviton-action}
    S_{\rm grav}=\frac{1}{64\pi}\int\dd^4x\,\qty(h_{ij}\Box h^{ij}+2\phi\,h_{ij}\delta_{\mu\nu}\partial^\mu\partial^\nu h^{ij}).
\end{equation}
Note that the interaction between the metric perturbation field and the Newtonian potential is of order $O(\phi h^2)$, while in Eq.~\eqref{Matter-action-2} they simultaneously couple with the system through an interaction of order $O(\phi h)$, which is therefore the dominant term. Furthermore, we remark that the interaction in Eq.~\eqref{Graviton-action}, upon quantisation of the radiation degrees of freedom, is physically associated with graviton scattering, for which the differential cross section is shown in Appendix~\ref{A:Gravitons-Newton}. This cross section is found to be strongly dominated by regions in which $\theta\ll1$, scaling as $\dv*{\sigma}{\Omega}\sim (G^2M_N^2/c^4)\theta^{-4}$, with $\theta$ being the angle between the directions of propagation of incident and scattered gravitons. This leads us to conclude that such an interaction is less dominant than the one that is coupled with the system. In fact, this interaction term is neglected in Ref.~\cite{Chawla2023}, where the authors compute quantum gravity corrections to the fall of test masses, for instance.

From now on, let us drop this interaction term and write the graviton action in Fourier space. We introduce the expansion in Fourier modes
\begin{equation} \label{Fourier-expansion-gravitons}
    h_{ij}(t,\vb{x})=\int\dd^3k\,\Tilde{h}_{ij}(t,\vb{k})e^{i\vb{k}\vdot\vb{x}}.
\end{equation}
The reality of $h_{ij}(t,\vb{x})$ implies that $\Tilde{h}^*(t,\vb{k})=\Tilde{h}(t,-\vb{k})$. We can take the Fourier transform of the metric perturbation field to be of the form
\begin{equation} \label{Amplitude(k)timesPolatizations}
    \Tilde{h}_{ij}(t,\vb{k})=\sum_{s=+,\cross}\epsilon_{ij}^s(\vb{k})q_s(t,\vb{k}),
\end{equation}
where $q_s(t,\vb{k})$ is the wave amplitude in Fourier space with polarisation $s$, and $\epsilon_{ij}^s(\vb{k})$ are the polarisation tensors satisfying the normalisation conditions $\epsilon_{ij}^s(\vb{k})\epsilon^{ij}_{s'}(\vb{k})=2\delta^s_{s'}$, the transversality $k^i\epsilon_{ij}^{s}(\vb{k})=0$ and the traceless conditions $\delta^{ij}\epsilon_{ij}^{s}(\vb{k})=0$.

Using the expansion~\eqref{Fourier-expansion-gravitons} the graviton action becomes
\begin{equation} \label{Graviton-action-Fourier-space-2}
    S_{\rm grav}=\int\dd t\int\dd^3k\,\sum_s\frac{m_{\rm g}}{2}\qty[\abs{\dot{q}_s(t,\vb{k})}^2-\vb{k}^2\abs{q_s(t,\vb{k})}^2],
\end{equation}
where we defined $m_{\rm g}=\pi^2/2$. The action~\eqref{Graviton-action-Fourier-space-2} has the form $S_{\rm grav}=\int\dd t\int\dd^3k\,\sum_sL_s(t,\vb{k})$, with Lagrangian $L_s(t,\vb{k})$ for each mode that describes a harmonic oscillator with mass $m_{\rm g}$ and frequency $\omega=\abs{\vb{k}}$. The associated Hamiltonian takes the usual form
\begin{equation} \label{Hamiltonian-individual-mode}
    H_s(t,\vb{k})=\frac{\abs{p_s(t,\vb{k})}^2}{2m_{\rm g}}+\frac{1}{2}m_{\rm g}\vb{k}^2\abs{q_s(t,\vb{k})}^2,
\end{equation}
with $p_s(t,\vb{k})=\pdv{L_s(t,\vb{k})}{\dot{q}_s(t,\vb{k})}=m_{\rm g}\dot{q}_s(t,\vb{k})$.

\subsection{The total action}

By introducing the Fourier expansion for the metric field, Eqs.~\eqref{Fourier-expansion-gravitons} and~\eqref{Amplitude(k)timesPolatizations}, the total action takes the form
\begin{equation} \label{Total-action}
    S_{\rm total}=S_{\rm sys}+S_{\rm grav}+\int\dd t\int\dd^3k\sum_sq_s(t,\vb{k})X^s(t,\vb{k}),
\end{equation}
where
\begin{align} \label{System-action}
    S_{\rm sys}&=\int\dd t\,\left\{ \frac{1}{2}m\delta_{ij}\dot{\xi}^i\dot{\xi}^j+\mathscr{L}(\varrho,\dot{\varrho}\Bar{t})\right. \nonumber \\
    &\left. -\frac{1}{2}\mathscr{L}(\varrho,\dot{\varrho}\Bar{t})\delta_{ij}\dot{\xi}^i\dot{\xi}^j-\frac{1}{2}L_{\rm rest}(\varrho,\dot{\varrho}\Bar{t})T_{ij}\xi^i\xi^j\right\} ,
\end{align}
$S_{\rm grav}$ is given by Eq. \eqref{Graviton-action-Fourier-space-2} and we have defined
\begin{align} \label{System-Xs-variable}
    X^s(t,\vb{k})&=-\frac{1}{4}\epsilon_{ij}^s(\vb{k})L_{\rm rest}(\varrho,\dot{\varrho}\Bar{t}){T^i}_k\xi^k\xi^j \nonumber \\
    &-\frac{1}{4}\epsilon_{ij}^s(\vb{k})\dv[2]{t}\qty[L_{\rm rest}(\varrho,\dot{\varrho}\Bar{t})\xi^i\xi^j].
\end{align}
Equation~\eqref{Total-action} is the classical action that describes the interaction between the various degrees of freedom of the system and the gravitational fields $\phi(\vb{x})$, a time-independent Newtonian potential, and $h_{\mu\nu}(t,\vb{x})$, a metric perturbation field propagating through spacetime. In the next section, we quantise both the system and the gravitational radiation field, the latter giving rise to the graviton bath.

\section{Time evolution of the reduced density matrix} 
\label{Sec:Time-evolution}

Let us now quantize the systems described in the last section. We will be interested in analysing quantum superposition states of the external DoFs of the composite particle, so both the internal DoFs and the gravitons will be treated as environments. Since they interact with each other, one cannot simply evaluate their separate contributions and then add everything together, as this would not take their coupling into account. We address this problem by using the Feynman-Vernon approach to open quantum systems~\cite{Feynman1963,Feynman2010,Calzetta2008}. First, we integrate the gravitational and internal DoFs variables. The entire procedure was described in~\cite{Moreira_2023}, so here we shall only outline the main steps and the modifications that arise by the introduction of the classical Newtonian potential.

Within the Feynman-Vernon approach to open quantum systems, the time evolution of the system reduced density matrix $\rho_{\rm sys}$, which is obtained from the total density matrix by taking a partial trace with respect to the environment variables, is given by
\begin{align}
    \rho_{\rm sys}(X,X',t)&=\int \dd X_0\dd X_0'\,\mathcal{J}_{\rm sys}(X,X',t|X_0,X_0',0) \nonumber \\
    &\times\rho_{\rm sys}(X_0,X_0',0),
\end{align}
where $\rho_{\rm sys}(X,X',t)=\mel{X}{\rho_{\rm sys}(t)}{X'}$, with $\ket{X}$ denoting a position eigenstate in the reduced Hilbert space of the system, and the evolution operator for the reduced density matrix reads
\begin{align}
    &\mathcal{J}_{\rm sys}(X,X',t|X_0,X_0',0) \nonumber \\
    &=\int\displaylimits_{\substack{X(0)\,=\,X_0 \\ X'(0)\,=\,X_0'}}^{\substack{X(t)\,=\,X \\ X'(t)\,=\,X'}}\mathcal{D}X\mathcal{D}X'\,e^{i\qty(S_{\rm sys}[X]-S_{\rm sys}[X'])}e^{iS_{\rm IF}[X,X',t]},
\end{align}
with $S_{\rm sys}$ denoting the part of the action that describes the system alone and $\mathcal{D}X$ denoting the integral measure of the path. Also, we assume that, at $t=0$, the system and the graviton bath were not correlated $\rho(0)=\rho_{\rm sys}(0)\otimes\rho_{\rm grav}(0)$. The functional $e^{iS_{\rm IF}[X,X',t]}$ is the influence functional, while $S_{\rm IF}$ represents the influence action.

In the case of Eq.~\eqref{Total-action}, the system is coupled with the infinite set of graviton modes, which are all independent of each other, as well as the two polarizations. They can thus be treated separately in such a way that the total influence action is the sum of the action corresponding to each mode and polarisation. Furthermore, we note that the environment action (the graviton action) is quadratic in the field amplitudes, and the coupling with the system variable $X^s(t,\vb{k})$, Eq.~\eqref{System-Xs-variable}, is linear. This is then a special case of the linear coupling model~\cite{Calzetta2008,Cho_2023,Hsiang_2024,Moreira_2023}, for which the influence action is found to be
\begin{widetext}
\begin{equation} \label{Grav-influence-action-2}
\begin{split}
    S_{\rm IF}[x,x']=\int \dd t\dd t'\left\{ \frac{1}{2}\qty[x_{ij}(t)-x'_{ij}(t)]D^{ijkl}_{\rm g}(t,t')\qty[x_{kl}(t')+x'_{kl}(t')] \right. \\
    \left. +\frac{i}{2}\qty[x_{ij}(t)-x'_{ij}(t)]N^{ijkl}_{\rm g}(t,t')\qty[x_{kl}(t')-x'_{kl}(t')]\right\},
\end{split}
\end{equation}
where we have defined the gravitational dissipation
\begin{subequations} \label{Grav-dissipation-and-noise-kernels}
\begin{equation} \label{Grav-dissipation-kernel}
\begin{split}
    D_{\rm g}^{ijkl}(t,t')&=\frac{1}{16}\sum_s\int\dd^3k\,\left[ \epsilon^{ij}(\vb{k})\epsilon^{kl}(\vb{k})\dv[2]{t}\dv[2]{{t'}}d_s(t,t',\vb{k})\right. \\
    &\hspace{0.3cm}\left. +\epsilon^{ij}(\vb{k})\epsilon^{nl}(\vb{k}){T_n}^k\dv[2]{t}d_s(t,t',\vb{k})+\epsilon^{nj}(\vb{k})\epsilon^{kl}(\vb{k}){T_n}^i\dv[2]{{t'}}d_s(t,t',\vb{k})\right] ,
\end{split}
\end{equation}
and noise kernels
\begin{equation} \label{Grav-noise-kernel}
\begin{split}
    N_{\rm g}^{ijkl}(t,t')&=\frac{1}{16}\sum_s\int\dd^3k\,\left[ \epsilon^{ij}(\vb{k})\epsilon^{kl}(\vb{k})\dv[2]{t}\dv[2]{{t'}}n_s(t,t',\vb{k})\right. \\
    &\hspace{0.3cm}\left. +\epsilon^{ij}(\vb{k})\epsilon^{nl}(\vb{k}){T_n}^k\dv[2]{t}n_s(t,t',\vb{k})+\epsilon^{nj}(\vb{k})\epsilon^{kl}(\vb{k}){T_n}^i\dv[2]{{t'}}n_s(t,t',\vb{k})\right] ,
\end{split}
\end{equation}
\end{subequations}
\end{widetext}
with $d_s(t,t',\vb{k})$ and $n_s(t,t',\vb{k})$ defined by
\begin{subequations} \label{Dissipation-and-noise-kernels}
\begin{equation} \label{Dissipation-kernel}
    d_s(t,t',\vb{k})=i\expval{\comm{q_s(t,\vb{k})}{q_s(t',\vb{k})}}_{\rm g}\theta(t-t'),
\end{equation}
and
\begin{equation} \label{Noise-kernel}
    n_s(t,t',\vb{k})=\frac{1}{2}\expval{\acomm{q_s(t,\vb{k})}{q_s(t',\vb{k})}}_{\rm g}.
\end{equation}
\end{subequations}
Also, we define
\begin{equation} \label{x-ij(t)-definition}
    x_{ij}(t)=L_{\rm rest}(\varrho,\dot{\varrho}\Bar{t})\xi_i(t)\xi_j(t).
\end{equation}
In Eqs.~\eqref{Dissipation-and-noise-kernels} the $q$'s stand for position operators in the Heisenberg picture, $\comm{\cdot}{\cdot}$ and $\acomm{\cdot}{\cdot}$ denote the commutator and anti-commutator of operators, respectively, while $\theta(x)$ is the Heaviside step function. The expectation values with the subscript 'g' are computed with respect to the initial state of the gravitons. Note that the influence action~\eqref{Grav-influence-action-2} has the same form as the one obtained in~\cite{Moreira_2023}. The difference now is that both the dissipation and noise kernels contain new terms that arise due to the presence of a Newtonian potential. This kind of modification of such kernels due to the presence of an external potential has been noticed in the context of the quantum Brownian motion, as in ref.~\cite{Colmenares2023}, for instance.

Just as is usually done, we can rewrite the term that contains the noise kernel by using a Gaussian stochastic variable $\mathcal{N}$~\cite{Cho2022,Moreira_2023}. This term will then be of order $O(\xi^2)$ while the one with the dissipation kernel will be of order $O(\xi^4)$, allowing us to approximate the influence functional by
\begin{equation} \label{Influence-functional-final-approximated}
    e^{iS_{\rm IF}[x,x']}\simeq\int\mathcal{D}\mathcal{N}\,\mathscr{P}[\mathcal{N}]\,e^{i\int \dd t\,\mathcal{N}^{ij}(t)\qty[x_{ij}(t)-x'_{ij}(t)]},
\end{equation}
in a perturbative approach. Here, $\mathscr{P}[\mathcal{N}]$ stands for a Gaussian probability density.

The influence functional~\eqref{Influence-functional-final-approximated} describes the influence of the graviton bath on the time evolution of the system, which is no longer unitary. Next, we can proceed to give the same treatment to the system internal DoFs. Following Ref.~\cite{Moreira_2023}, we assume that the external and internal DoFs are initially uncorrelated, and we take an internal Lagrangian of the form
\begin{equation}
    \mathscr{L}(\varrho,\dot{\varrho}\,\bar{t})\simeq\sum_\alpha\qty(\frac{1}{2}\mu_\alpha\dot{\varrho}_\alpha^2-\vartheta_\alpha\varrho_\alpha-\frac{1}{2}\mu_\alpha\varpi_\alpha^2\varrho_\alpha^2),
\end{equation}
where $\mu_{\alpha}$, $\vartheta_{\alpha}$ and $\varpi_{\alpha}$ are constants. This leads, once again, to the linear coupling model, and one obtains another influence action in terms of (internal DoFs) dissipation and noise kernels. Here also, one can argue that the dominant contribution comes from noise, and the external DoFs density matrix is found to be given by
\begin{align} \label{External-DoFs-density-matrix-0}
    &\rho_{\rm ext}(\xi,\xi',t)=\int \dd\xi(0)\dd\xi'(0)\,\rho_{\rm ext}(\xi(0),\xi'(0),0) \nonumber \\
    &\times\int\mathcal{D}\xi\mathcal{D}\xi'\,e^{\frac{i}{2}m\delta^{ij}\int \dd t\,\qty(\dot{\xi}_i\dot{\xi}_j-\dot{\xi}_i'\dot{\xi}_j')} \nonumber \\
    &\times\int\mathcal{D}\mathcal{N}\,\mathscr{P}[\mathcal{N}]\,e^{-im\int \dd t\,\mathcal{N}^{ij}\qty(\xi_i\xi_j-\xi_i'\xi_j')} \nonumber \\
    &\times e^{-\frac{1}{2}\int \dd t\dd t'\,\qty[Y(t)-Y'(t)]N_{\rm int}(t,t')\qty[Y(t')-Y'(t')]},
\end{align}
where
\begin{equation}
    Y(t)=\frac{1}{2}\delta_{ij}\dot{\xi}^i\dot{\xi}^j-\mathcal{N}_{ij}\xi^i\xi^j-1
\end{equation}
and
\begin{equation} \label{Internal-dofs-noise-kernel}
    N_{\rm int}(t,t')=\frac{1}{2}\sum_\alpha\vartheta_\alpha^2\expval{\acomm{\varrho_\alpha(t)}{\varrho_\alpha(t')}}_{\rm int}
\end{equation}
is the internal DoFs noise kernels. Now, the $\varrho$'s are operators in the Heisenberg picture, and expectation values with the subscript 'int' are computed with respect to the initial state of the internal DoFs. Also, in arriving at Eq.~\eqref{External-DoFs-density-matrix-0} we made a change of variables such that the stochastic variable $\mathcal{N}_{ij}$ satisfies
\begin{subequations}  \label{New-Stochastic-averages}
\begin{equation}
    \expval{\mathcal{N}^{ij}(t)}_{\rm sto}=-\frac{1}{2}T_{ij}
\end{equation}
and
\begin{equation}
    \expval{\mathcal{N}^{ij}(t)\mathcal{N}^{kl}(t')}_{\rm sto}=N_{\rm g}^{ijkl}(t,t')+O(T_{ij}^2),
\end{equation}
\end{subequations}
where the averages are computed with the Gaussian probability density. These stochastic averages can be employed, provided that we work on a perturbative regime, in order to obtain the external degrees of freedom density matrix as
\begin{widetext}
\begin{equation} \label{External-DoFs-density-matrix}
\begin{split}
    &\rho_{\rm ext}(\xi,\xi',t)=\int \dd\xi(0)\dd\xi'(0)\,\rho_{\rm ext}(\xi(0),\xi'(0),0) \int\mathcal{D}\xi\mathcal{D}\xi'\,e^{i\qty(S_{\rm ext}[\xi]-S_{\rm ext}[\xi'])} \\
    &\hspace{0.3cm}\times\exp\qty{-\frac{1}{4}\int\dd t\dd t'\,\qty[\frac{1}{2}\delta^{ij}\delta^{kl}y_{ij}(t)N_{\rm int}(t,t')y_{kl}(t')+\delta^{ij}T^{kl}y_{ij}(t)N_{\rm int}(t,t')w_{kl}(t')]} \\
    &\hspace{0.3cm}\times\exp\qty{-\int\dd t\dd t'\,w_{ij}(t)\qty[\frac{m^2}{2}+N_{\rm int}(t,t')]N_{\rm g}^{ijkl}(t,t')w_{kl}(t')},
\end{split}
\end{equation}
\end{widetext}
where
\begin{equation}
    S_{\rm ext}[\xi]=\frac{m}{2}\int\dd t\,\qty(\delta^{ij}\dot{\xi}_i\dot{\xi}_j+T^{ij}\xi_i\xi_j),
\end{equation}
and we have defined
\begin{subequations}
\begin{equation}
    w_{ij}(t)=\xi_i(t)\xi_j(t)-\xi_i'(t)\xi_j'(t)
\end{equation}
and
\begin{equation}
    y_{ij}(t)=\dot{\xi}_i(t)\dot{\xi}_j(t)-\dot{\xi}_i'(t)\dot{\xi}_j'(t).
\end{equation}
\end{subequations}

Equation~\eqref{External-DoFs-density-matrix} is the main result of this section. It encapsulates how the graviton bath and the system own internal structure affect the time evolution of external centre-of-mass variables in an effective non-unitary manner. Note that these contributions affect only non-diagonal elements in the position representation of the density matrix, and we will see how they lead to decoherence of quantum superpositions of the external DoFs.

\section{The decoherence function} \label{Sec:Decoherence-function}

From now on, let us consider a special case in which the system can move only along two classically distinguishable paths $\xi^{(1)}(t)$ and $\xi^{(2)}(t)$. Then, Eq.~\eqref{External-DoFs-density-matrix} can be written as
\begin{align}
    &\rho_{\rm ext}(\xi,\xi',t)=\int\dd\xi_0\dd\xi_0'\rho_{\rm ext}(\xi_0,\xi_0',0) \nonumber \\
    &\times\sum_{m,n=1}^2e^{i\qty{S_{\rm ext}\qty[\xi^{(m)}]-S_{\rm ext}\qty[\xi^{(n)}]}} \nonumber \\
    &\hspace{0.3cm}\times \eval{e^{-\Gamma\qty[\xi^{(m)},\xi^{(n)},t]}}_{\xi^{(m)}(0)=\xi_0,\,\,\xi^{(n)}(0)=\xi_0'}^{\xi^{(m)}(t)=\xi,\,\,\xi^{(n)}(t)=\xi'},
\end{align}
where we introduced the functional $\Gamma\qty[\xi^{(m)},\xi^{(n)},t]$ whose explicit form can be read from Eq.~\eqref{External-DoFs-density-matrix} setting $\xi\to\xi^{(m)}$ and $\xi'\to\xi^{(n)}$. This is a functional of the paths $\xi^{(1)}(t)$ and $\xi^{(2)}(t)$ that satisfies
\begin{subequations} \label{Properties-decoherence-functional}
\begin{equation}
    \Gamma\qty[\xi^{(m)},\xi^{(n)},t]=\Gamma\qty[\xi^{(n)},\xi^{(m)},t],
\end{equation}
and
\begin{equation}
    \Gamma\qty[\xi^{(m)},\xi^{(m)},t]=0.
\end{equation}
\end{subequations}
Due to these properties, we only need to consider $\Gamma\qty[\xi^{(1)},\xi^{(2)},t]\equiv\Gamma(t)$. When $\Gamma(t)\geq0$, we call it the \emph{decoherence function}, since in that case it leads to an exponential decay of the off-diagonal elements of the density matrix (the coherences).

For simplicity, let us take $\xi_i^{(m)}(t)=\xi^{(m)}(t)\delta_{i3}$, that is, unidimensional paths in the $z-$direction. Now, consider a scheme in which both paths start at the same point $\xi_0=\xi_0'$ and end at another point $\xi=\xi'$. This is the typical scenario in which an initially localised quantum system undergoes a superposition of paths and is then recombined after time $t_f$ in order for its interference patterns to be analysed.

First, let us define the variables
\begin{subequations} \label{Configurations-of-paths}
\begin{equation}
    \Xi(t)\equiv\frac{1}{2}\qty[\xi^{(1)}(t)+\xi^{(2)}(t)],\hspace{0.3cm}\Delta\xi(t)\equiv\xi^{(1)}(t)-\xi^{(2)}(t),
\end{equation}
and
\begin{equation}
    V(t)\equiv\dv{t}\Xi(t),\hspace{0.5cm}\Delta v(t)\equiv\dv{t}\Delta\xi(t).
\end{equation}
\end{subequations}
The noise kernel $N_{\rm int}(t,t')$ was computed in~\cite{Moreira_2023} considering the internal DoFs to represent an Ohmic bath described by the coupling constant $\eta$ in thermal equilibrium at high temperature $T_{\rm int}\gg\abs{t-t'}^{-1}$, resulting in
\begin{equation}
    N_{\rm int}(t,t')=\eta\pi T_{\rm int}\delta(t-t').
\end{equation}
The Ohmic high-temperature approximation should be understood as an effective description valid when the internal correlation time is much shorter than the characteristic timescale of the external centre-of-mass dynamics, so that the internal noise can be treated as approximately Markovian. It also assumes that the internal degrees of freedom form a sufficiently dense set of modes with an approximately Ohmic spectral density over the frequency range probed by the external motion. These assumptions are useful because they lead to the local kernel in the above equation, allowing us to obtain explicit analytic expressions for the decoherence function and to focus on the interplay between the internal dynamics, the graviton bath and the Newtonian potential. More general internal environments could be considered by replacing $N_{\mathrm{int}}(t,t')$ by the appropriate noise kernel. For instance, sub-Ohmic or super-Ohmic spectral densities, low-temperature internal states, or a finite number of discrete internal modes would generally produce nonlocal kernels and could modify the detailed time dependence of the decoherence function. The systematic study of such alternatives is left for future work.

The decoherence function can then be written as
\begin{align} \label{Decoherence-function-general-configuration}
    \Gamma(t_f)&=\frac{1}{2}\eta\pi T_{\rm int}\int_0^{t_f}\dd t\,\qty[V(t)\Delta v(t)]^2 \nonumber \\
    &+\eta\pi T_{\rm int}T_{zz}\int_0^{t_f}\dd t\,V(t)\Delta v(t)\Xi(t)\Delta\xi(t) \nonumber \\
    &+2m^2\int_0^{t_f}\dd t\dd t'\,\Xi(t)\Delta\xi(t)N_{\rm g}(t,t')\Xi(t')\Delta\xi(t') \nonumber \\
    &+4\eta\pi T_{\rm int}\int_0^{t_f}\dd t\,\qty[\Xi(t)\Delta\xi(t)]^2N_{\rm g}(t),
\end{align}
where we denote $N_{\rm g}(t,t')\equiv N_{\rm g}^{3333}(t,t')$ and $N_{\rm g}(t)\equiv\lim_{t'\to t}N_{\rm g}(t,t')$.

Now we can consider specific configurations of the superposition state. Our main choice, which we shall refer to as \emph{Configuration 1}, is described by~\cite{Kanno2021}
\begin{subequations} \label{Configuration1}
\begin{equation}
    \Xi(t)=\Xi=\textrm{constant in time}
\end{equation}
and
\begin{equation}
    \Delta\xi(t)=\left\{
    \begin{array}{ll}
        2vt &\textrm{for}\hspace{0.2cm}0<t\leq t_f/2\\
        2v(t_f-t) &\textrm{for}\hspace{0.2cm}t_f/2<t<t_f
    \end{array}
    \right.,
\end{equation}
for some constant velocity $v$. Note that this implies $V(t)=\dv*{\Xi}{t}=0$.
\end{subequations}

For Configuration 1, the decoherence rate becomes
\begin{align} \label{Dec-rate-Conf-1-general}
    \Gamma_1(t_f)&=8m^2\Xi^2v^2\left[ \int_0^{t_f/2}\dd t\dd t'\,tt'N_{\rm g}(t,t')\right. \nonumber \\
    &\hspace{0.5cm}+\int_{t_f/2}^{t_f}\dd t\dd t'\,(t_f-t)(t_f-t')N_{\rm g}(t,t') \nonumber \\
    &\hspace{0.5cm}\left. +2\int_0^{t_f/2}\dd t\int_{t_f/2}^{t_f}\dd t'\,t(t_f-t')N_{\rm g}(t,t')\right] \nonumber \\
    &+16\eta\pi T_{\rm int}\Xi^2v^2\left[ \int_0^{t_f/2}\dd t\,t^2N_{\rm g}(t)\right. \nonumber \\
    &\hspace{0.5cm}\left. +\int_{t_f/2}^{t_f}\dd t\,(t_f-t)^2N_{\rm g}(t)\right] .
\end{align}
This is the same expression we obtained in~\cite{Moreira_2023}, except that now there are corrections to the gravitational noise kernel coming from the Newtonian potential. Note from Eq.~\eqref{Dec-rate-Conf-1-general} that within this configuration, all decoherence comes either from the gravitons alone or their interplay with the internal DoFs. If there were no graviton bath, we would have $\Gamma_1(t)=0$. This is not generally true, as we can see by considering a different configuration for the superposition state.

For illustrative purposes, let us take a second configuration (\emph{Configuration 2}), which is described by linear paths with different constant speeds, $\xi^{(m)}=v_mt$. In that case, we have
\begin{subequations}
\begin{equation}
    \Xi(t)=Vt,\hspace{0.5cm}\Delta\xi(t)=\Delta vt,
\end{equation}
and
\begin{equation}
    V=\frac{v_1+v_2}{2},\hspace{0.5cm}\Delta v=v_1-v_2.
\end{equation}
\end{subequations}
This could be relevant for systems in a superposition state such that its components travel at different speeds, like mass superpositions whose wave packets propagate with different group velocities.

For Configuration 2, the decoherence rate becomes
\begin{align} \label{Dec-rate-Conf-2-general}
    \Gamma_2(t_f)&=\frac{1}{4}\eta\pi T_{\rm int}\qty(v_1^2-v_2^2)^2 \nonumber \\
    &\hspace{0.5cm}\times\qty[\frac{t_f}{2}+\frac{T_{zz}t_f^3}{3}+4\int_0^{t_f}\dd t\,t^4N_{\rm g}(t)] \nonumber \\
    &+\frac{m^2}{2}\qty(v_1^2-v_2^2)^2\int_0^{t_f}\dd t\dd t'\,(tt')^2N_{\rm g}(t,t').
\end{align}
Note that if we were to turn off the graviton bath, one would still have decoherence due to the coupling between internal and external DoFs mediated by special relativistic effects and the Newtonian potential (this is the kind of gravitational decoherence explored in~\cite{Pikovski2015}, for instance). Additionally, one can also see that while $\Gamma_1(t)$ scales as $v^2$, $\Gamma_2(t)$ depends on $v^4$, making the former more relevant to nonrelativistic systems such as the ones we consider here.

It is now only a matter of a long but straightforward calculation to work out the various explicit expressions for the decoherence function considering both configurations and gravitons to be in the four distinct initial states we considered in~\cite{Moreira_2023}. We compute the gravitational noise kernel for vacuum, thermal, coherent, and squeezed initial states in the Appendix~\ref{A:Gravitational-Noise-Kernel}. Additionally, let us recall from Eq.~\eqref{Tidal-tensor} that $T_{zz}=2M_N/R_N^3$. Also, it will be interesting to return the universal constants $\hbar$, $c$, $G$ and $k_B$, so that we present the results in SI units rather than Planck units. The explicit results involve rather long expressions, so we list them in the Appendix~\ref{App:Explicit-expressions-dec-func} for both configurations. However, since the relevant physics of graviton induced decoherence is expected to be independent of our choice, we will stick with Configuration 1 in the remainder of this work.

\subsection{Vacuum state} \label{Sec:Vacuum-state}

\begin{figure}[!ht]
    \centering
    \includegraphics[width=\linewidth]{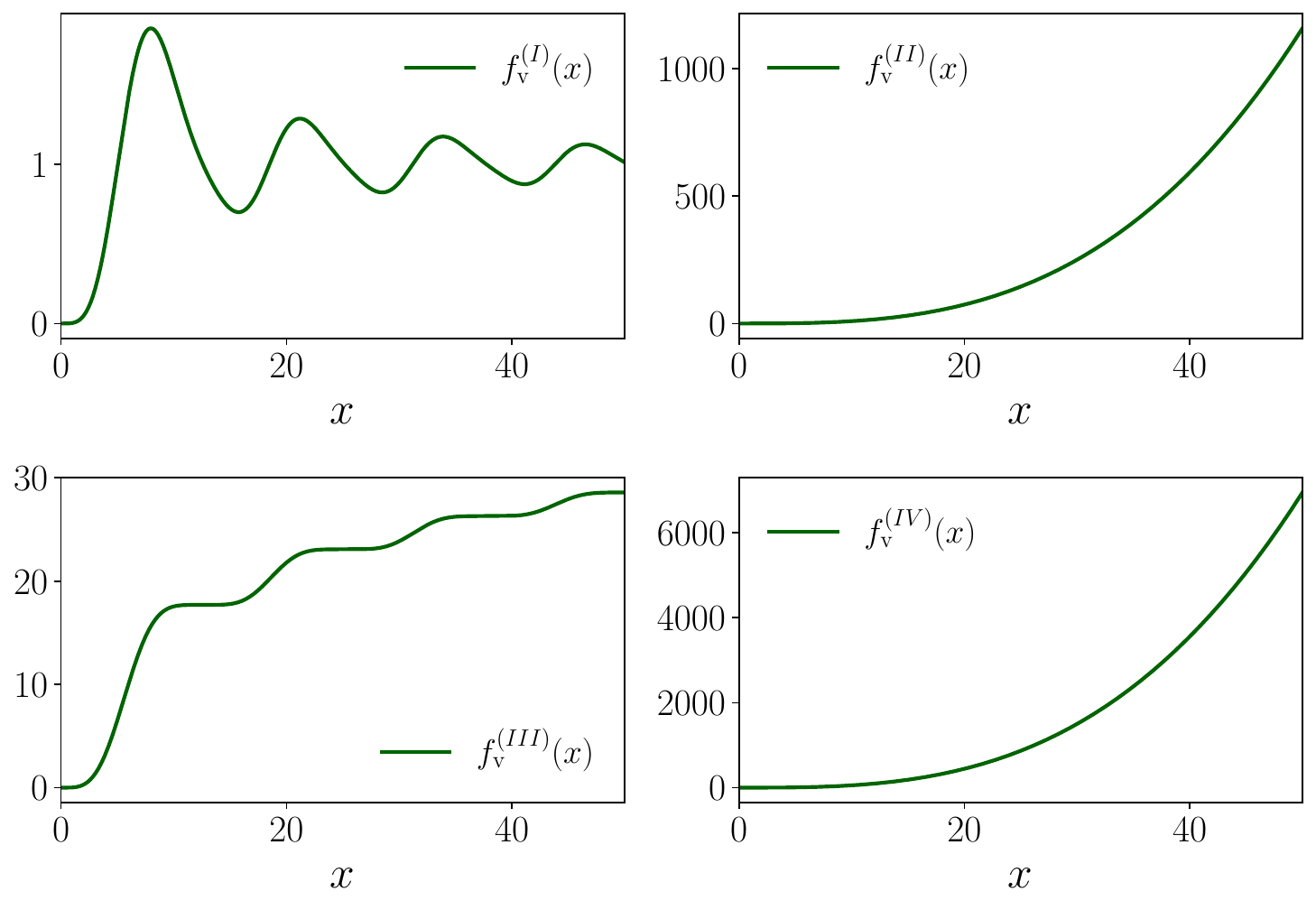}
    \caption{\justifying{Different contributions for the decoherence function considering the gravitons to be initially in the vacuum state.}}
    \label{Fig:vacuum_state}
\end{figure}

For gravitons initially in the vacuum state, the decoherence function (for Configuration 1) reads
\begin{align} \label{Vacuum-decoherence-function-Conf-1}
    &\Gamma_1^{(\textrm{v})}(t)=\frac{16\Xi^2v^2m^2}{5\pi E_{\rm P}^2} \nonumber \\
    &\times\left\{ \qty(\frac{\Lambda}{\hbar})^2\qty[f_{\rm v}^{(I)}\qty(\frac{\Lambda t}{\hbar})+\frac{\kappa}{E_{\rm rest}^2}f_{\rm v}^{(II)}\qty(\frac{\Lambda t}{\hbar})]\right. \nonumber \\
    &\hspace{0.5cm}\left. -\frac{GM_N}{R_N^3}\qty[f_{\rm v}^{(III)}\qty(\frac{\Lambda t}{\hbar})+\frac{\kappa}{E_{\rm rest}^2}f_{\rm v}^{(IV)}\qty(\frac{\Lambda t}{\hbar})]\right\} ,
\end{align}
where $\kappa=\eta\pi k_BT_{\rm int}\Lambda$, $E_{\rm rest}=mc^2$ and $E_{\rm P}=\sqrt{\hbar c^5/G}\simeq2.0\times10^9\,\textrm{J}$ is the Planck energy. The explicit expressions for the functions $f_{\rm v}(x)$ are given in Appendix~\ref{App:Explicit-expressions-dec-func}, while some of their relevant properties are shown in Table~\ref{tab:vacuum_state} (see also Figure~\ref{Fig:vacuum_state}).
\begin{table*}[!ht]
\centering

\begin{tabular}{llcc}
\toprule

\multicolumn{4}{c}{\textbf{Vacuum state}} \\
\cmidrule(lr){1-4}

\textbf{Contribution} & \textbf{Function} & \textbf{Behavior for $x \ll 1$} & \textbf{Behavior for $x \gg 1$} \\
\midrule

G     & $f_{\rm v}^{(I)}(x)$   & $\frac{x^4}{288}+O(x^5)$          & $1+O(1/x)$ \\
G+I   & $f_{\rm v}^{(II)}(x)$  & $\frac{x^3}{108}$                 & $\frac{x^3}{108}$ \\
G+N   & $f_{\rm v}^{(III)}(x)$ & $\frac{x^4}{48}+O(x^5)$           & $8\gamma_E-\frac{32}{3}\ln2+8\ln x+O(1/x)$ \\
G+N+I & $f_{\rm v}^{(IV)}(x)$  & $\frac{x^3}{18}$                  & $\frac{x^3}{18}$ \\

\bottomrule
\end{tabular}

\caption{\justifying{Different contributions for the decoherence function considering the gravitons to be initially in the \textbf{vacuum state}. On the "Contribution" column, "G" means gravitons, "I" means internal DoFs, while "N" stands for Newtonian potential.}}
\label{tab:vacuum_state}

\end{table*}

The contributions coming from the interplay between the gravitons and the internal DoFs of the system are scaled by the ratio
\begin{equation} \label{Ratio-int}
    \mathscr{R}=\frac{\kappa}{E_{\rm rest}^2}=\frac{\eta\pi k_BT_{\rm int}\Lambda}{m^2c^4},
\end{equation}
which depends on the mass of the composite system, the dimensionless coupling between the external and internal DoFs, $\eta$, as well as on the internal temperature, which can be of the order $T_{\rm int}\sim10^4$ K for complex molecules~\cite{Hornberger2012}. The other parameter is the graviton energy cutoff $\Lambda$. As we discussed in Sec.~\ref{Sec:Classical-action}, this arises not due to some limiting validity of the theory in a fundamental energy scale, as UV cutoffs usually appear in quantum field theories, but rather due to detector limitations. To be more precise, the geodesic deviation between the reference mass $M$ and the composite particle with mass $m$ is not expected to be sensitive to incident gravitational waves whose wavelengths are much smaller than some typical initial separation. Incident waves with energy much higher than $\hbar c/L_0$, with $L_0$ denoting some typical initial geodesic separation (the detector size), cannot influence the system in any way. We then take the energy cutoff to be precisely this bound, $\Lambda\sim\hbar c/L_0$, as in~\cite{Kanno2021,Cho2022}.

We can rewrite the ratio~\eqref{Ratio-int} as
\begin{equation} \label{Ratio-int-2}
    \mathscr{R}=\eta\frac{L(m,T_{\rm int})}{L_0},
\end{equation}
where $L(m,T_{\rm int})$ is a typical length scale determined by the properties of the system. For example, we can take $T_{\rm int}\sim10^4$ K and $m\sim10^{-22}$ kg, for which case we find $L(m,T_{\rm int})\sim L_{\rm P}\simeq1.6\times10^{-35}\,\textrm{m}$, with $L_{\rm P}$ denoting Planck length. Hence, although the ratio $\mathscr{R}$ can be increased by increasing the coupling between external and internal DoFs, and decreasing the detector size, one can expect to have $\mathscr{R}\ll1$ for typical systems. However, this does not mean that one can simply drop such contributions, as their relevance must be analysed by looking at the behaviours of the functions listed in Table~\ref{tab:vacuum_state}.

\emph{Short-time limit}. Let us begin by considering short times, $t\ll\hbar/\Lambda$, for which we can use the expansions shown in Table~\ref{tab:vacuum_state} and write the decoherence function as
\begin{align}
    \Gamma_1^{(\textrm{v})}(t)&=\frac{8\Xi^2v^2m^2}{15\pi E_{\rm P}^2}\qty[\frac{1}{6}\qty(\frac{\Lambda}{\hbar})^2-\frac{GM_N}{R_N^3}] \nonumber \\
    &\times\qty(\frac{\Lambda t}{\hbar})^3\qty(\frac{1}{8}\frac{\Lambda t}{\hbar}+\frac{1}{3}\frac{\kappa}{E_{\rm rest}^2}).
\end{align}

First, we note that there seems to be a competing behaviour between the squared frequencies $\qty(\Lambda/\hbar)^2$ and $GM_N/R_N^3$, and decoherence occurs only if the former is greater than the latter. Typically, this happens to be the case, as one can see from Table~\ref{tab:table-comparing-squared-frequencies}, where we estimate some values of both squared frequencies for different detector sizes and different sources of Newtonian gravitational potential. However, it seems to be possible, at least in principle, to have a situation in which the reversed scenario holds by increasing the detector size as well as the density of the Newtonian source, although one must be careful not to violate any of our assumptions in doing so. For instance, we show the result of the tidal squared frequency for neutron stars for informational purposes, but let us not forget that we are working in a perturbative regime, and dropping higher order terms in the tidal tensor. We will return to this issue in Sec.~\ref{Grav-rec}.
\begin{table*}[!ht]
\centering

\begin{minipage}[t]{0.45\textwidth}
    \centering
    \begin{tabular}{ccc}
    \toprule
    \multicolumn{3}{c}{\textbf{Estimating $\qty(\Lambda/\hbar)^2$}} \\
    \cmidrule(r){1-3}
    $L_0$ [m] & $\qty(\Lambda/\hbar)^2$ [$\textrm{s}^{-2}$] & $t_{\rm max}=\hbar/\Lambda$ [s] \\
    \midrule
    $10^{-6}$ & $9.0\times10^{28}$ & $3.3\times10^{-15}$ \\
    $10^{3}$ & $9.0\times10^{10}$ & $3.3\times10^{-6}$ \\
    $10^{9}$ & $9.0\times10^{-2}$ & $3.3$ \\
    \bottomrule
    \end{tabular}
\end{minipage}
%
\hspace{0.05\textwidth} 
%
\begin{minipage}[t]{0.35\textwidth} 
    \centering
    \begin{tabular}{cc}
    \toprule
    \multicolumn{2}{c}{\textbf{Estimating $GM_N/R_N^3$}} \\
    \cmidrule(l){1-2}
    Source & $GM_N/R_N^3$ [$\textrm{s}^{-2}$] \\
    \midrule
    Sun & $3.9\times10^{-7}$ \\
    Earth & $1.5\times10^{-6}$ \\
    Neutron star & $1.7\times10^{8}$ \\
    \bottomrule
    \end{tabular}
\end{minipage}

\caption{\justifying{Typical squared frequencies. On the left we estimate the magnitude of $\qty(\Lambda/\hbar)^2$ for some detector sizes $L_0$, together with the maximum time $t_{\rm max}$ establishing the validity of the short-time approximation. On the right we estimate the magnitude of $GM_N/R_N^3$ for some sources of the Newtonian potential.}}
\label{tab:table-comparing-squared-frequencies}

\end{table*}

Now, recall that, while $L_0$ represents a typical geodesic deviation with respect to the much more massive mass $M$, $\Xi$ denotes the size of the superposition in the reference frame centred on it. For simplicity, we can consider that one of the paths is close enough to the reference mass $M$ such that $\Xi\sim L_0$, leading to
\begin{equation} \label{dec-func-short-time}
    \Gamma_1^{(\textrm{v})}(t)=\frac{4}{45\pi}\delta\Omega\qty(\frac{v}{c})^2\qty(\frac{m}{M_{\rm P}})^2\qty(\frac{\Lambda t}{\hbar})^3\qty(\frac{1}{8}\frac{\Lambda t}{\hbar}+\frac{1}{3}\frac{\kappa}{E_{\rm rest}^2}),
\end{equation}
where $M_{\textrm{P}}=E_{\rm P}/c^2\simeq2.2\times10^{-8}$ kg is the Planck mass, and we define
\begin{equation} \label{delta-Omega}
    \delta\Omega\equiv1-6\qty(\frac{\hbar}{\Lambda})^2\frac{GM_N}{R_N^3}.
\end{equation}
Note that $\delta\Omega\leq1$ and, typically, $\delta\Omega\simeq1$.

Next, we note from Eq.~\eqref{dec-func-short-time} that, as long as $t\gg(8/3)\eta t(m,T_{\rm int})$, with $t(m,T_{\rm int})=L(m,T_{\rm int})/c$ denoting some typical timescale determined by the system, the contribution from the interplay between the gravitons and the internal DoFs becomes negligible, i.e., in the regime in which
\begin{equation}
    \frac{8}{3}\eta\frac{L(m,T_{\rm int})}{c}\ll t\ll\frac{L_0}{c},
\end{equation}
the interaction with the graviton bath alone dominates, and the decoherence function becomes
\begin{equation} \label{Final-dec-func-short-time}
    \Gamma_1^{(\textrm{v})}(t)=\frac{1}{90\pi}\delta\Omega\qty(\frac{v}{c})^2\qty(\frac{m}{M_{\rm P}})^2\qty(\frac{\Lambda t}{\hbar})^4.
\end{equation}
The decoherence time $t_{\rm dec}$, defined by the condition $\Gamma_1^{(\textrm{v})}(t_{\rm dec})=1$, is found to be given by\footnote{Here we denote $t_{\rm dec}$ for decoherence times computed in the short-time limit, while $\tau_{\rm dec}$ denotes decoherence times computed in the long-time limit.}
\begin{equation} \label{Short-time-dec-time-vacuum}
    t^{(\textrm{v})}_{\rm dec}=\frac{\hbar}{\Lambda}\sqrt{\sqrt{\frac{90\pi}{\delta\Omega}}\frac{c}{v}\frac{M_{\rm P}}{m}}.
\end{equation}
Since we are considering a regime in which $t^{(\textrm{v})}_{\rm dec}\ll\hbar/\Lambda$, we are led to the conclusion that graviton induced decoherence happens for systems that satisfy
\begin{equation} \label{Short-time-decoherence-condition}
    mv\gg\sqrt{\frac{90\pi}{\delta\Omega}}M_{\rm P}c,
\end{equation}
which is qualitatively the same result found in~\cite{Kanno2021}. For $\delta\Omega\simeq1$, this means that we must have $mv\gg110\,\textrm{kg}\cdot\textrm{m}/\textrm{s}$.

Equation~\eqref{Short-time-decoherence-condition} suggests that the observation of graviton induced decoherence requires the preparation of spatial quantum superpositions of macroscopic masses. Even if we were to extrapolate our results to the ultrarelativistic limit $v\simeq c$, the decoherence condition would require masses that satisfy $m\gg3.7\times10^{-7}$ kg. In conclusion, the short-time limit is dominated solely by the graviton bath, and decoherence does not occur for microscopic masses. We can hope that the scenario will be different for a long time. However, if we consider only the purely graviton contribution, one can see from Table~\ref{tab:vacuum_state} that $f_{\rm v}^{(I)}$ tends to a constant value as $t\to\infty$, and the decoherence function saturates at
\begin{equation} \label{Gamma_sat}
    \Gamma^{(\textrm{v})}_{1,\rm sat}=\frac{16}{5\pi}\delta\Omega\qty(\frac{v}{c})^2\qty(\frac{m}{M_{\rm P}})^2,
\end{equation}
increasing no further, meaning that the graviton bath alone cannot decohere spatial superpositions of a microscopic mass. However, if we consider the graviton interplay with the other contributions (internal DoFs and Newtonian potential), the scenario is different, since the other functions exhibit no such behaviour as $t\to\infty$. Let us then turn our attention to Eq.~\eqref{Vacuum-decoherence-function-Conf-1} and consider the long-time limit.

\emph{Long-time limit}. Table~\ref{tab:vacuum_state} shows that, for $x\gg1$, the functions $f_{\rm v}^{(II)}(x)$ and $f_{\rm v}^{(IV)}(x)$, which include the contributions of internal DoFs, scale as $x^3$, while $f_{\rm v}^{(I)}(x)$ remains constant and $f_{\rm v}^{(III)}(x)$ grows in a much slower logarithmic rate. This suggests that despite the ratio $\mathscr{R}$ being typically very small, the G+I (graviton+internal DoFs) contributions will dominate over the pure graviton ones for sufficiently long times, and decoherence will eventually become inevitable.

In the long-time limit, using the definition~\eqref{delta-Omega} and taking $\Xi\sim L_0$, Eq.~\eqref{Vacuum-decoherence-function-Conf-1} becomes
\begin{equation} \label{Final-dec-func-long-time}
    \Gamma_1^{(\textrm{v})}(t)=\frac{4}{135}\delta\Omega\qty(\frac{v}{c})^2\qty(\frac{\eta k_BT_{\rm int}\Lambda}{E_{\rm P}^2})\qty(\frac{\Lambda t}{\hbar})^3.
\end{equation}

The first thing we note is that the decoherence function~\eqref{Final-dec-func-long-time} does not seem to contain any explicit dependence on the mass of the composite particle, contrary to the usual models of gravitational decoherence. However, it should be noted that the long-term limit was established according to the ratio $\kappa/E_{\rm rest}^2$, which \emph{depends} on the mass. Additionally, explicit dependence on $m$ could arise through the dimensionless coupling constant $\eta$.

From Eq.~\eqref{Final-dec-func-long-time} we can compute the decoherence time, for which $\Gamma_1^{(\textrm{v})}(\tau_{\rm dec})=1$. One finds
\begin{align} \label{Long-time-dec-time-vacuum}
    \tau_{\rm dec}^{(\textrm{v})}&=\qty[\frac{135}{4}\frac{1}{\delta\Omega}\qty(\frac{c}{v})^2\frac{\hbar^4c^5}{Gk_B}\frac{1}{\eta T_{\rm int}\Lambda^4}]^{1/3} \nonumber \\
    &=\qty[\frac{135}{4}\frac{1}{\delta\Omega}\qty(\frac{c}{v})^2\frac{cL_0^4}{Gk_B\eta T_{\rm int}}]^{1/3}
\end{align}
In order to provide a numerical estimate, let us consider a strong coupling scenario and take $\eta\sim1$. For the internal temperature, we consider $T_{\rm int}\sim10^4$ K, $v\sim10^{-6}c$ as a typical molecular speed, $L_0\sim10^{-9}$ m, and $\delta\Omega\sim1$. We then obtain $\tau_{\rm dec}^{(\textrm{v})}\sim10^5$ s (about $1.2$ days), which is infinitely long compared to typical decoherence times. Nevertheless, this result shows that decoherence eventually does happen when we consider the interplay between the gravitons and the internal DoFs of the system, as opposed to the case where only the graviton bath contributes directly. Finally, we recall that this decoherence time was obtained by considering the initial vacuum state for the gravitons, and we will now proceed to show that such a timescale can decrease significantly for other initial states.

\subsection{Thermal, coherent and squeezed states}

\begin{figure}[!ht]
    \centering
    \includegraphics[width=\linewidth]{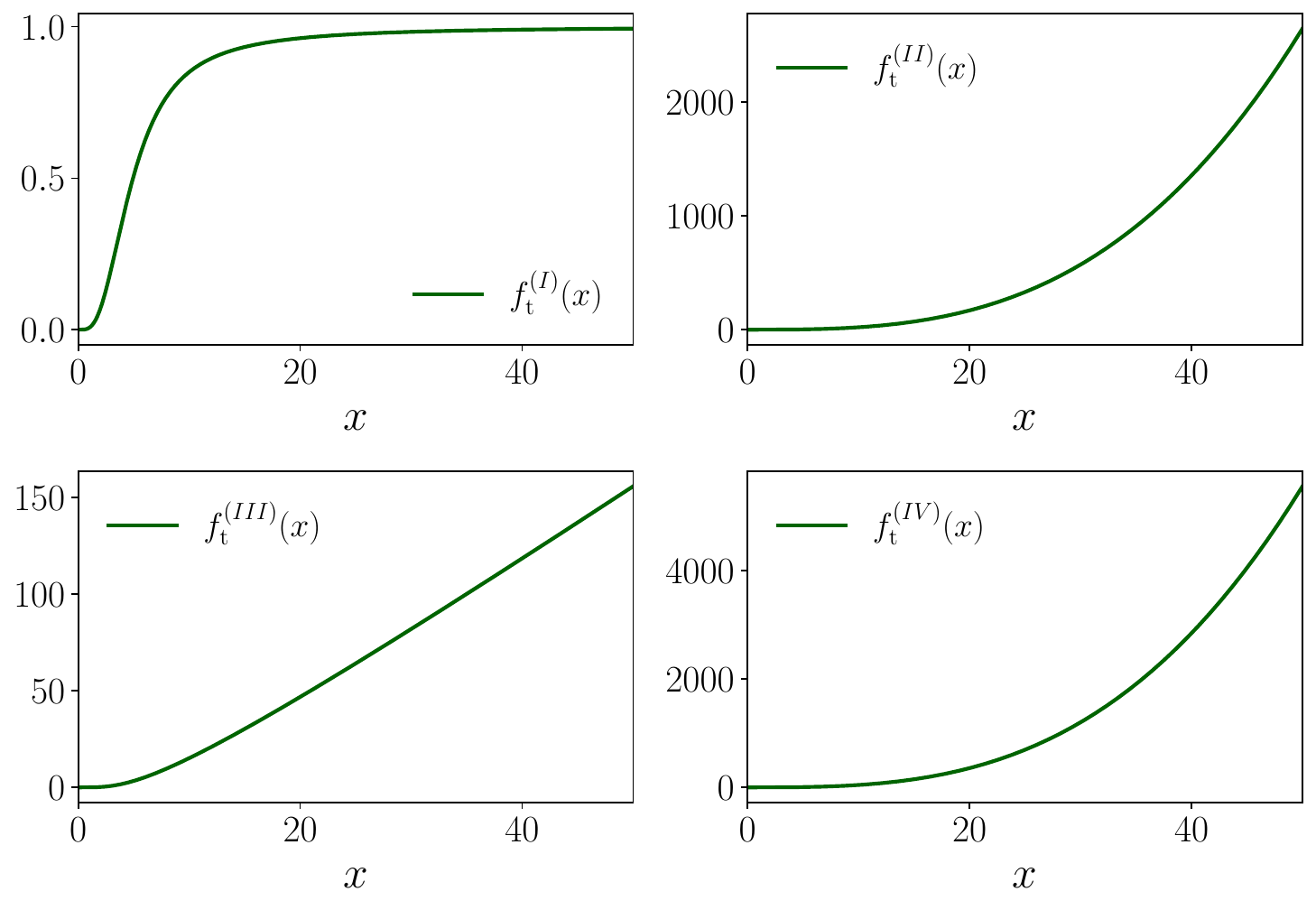}
    \caption{\justifying{Different contributions for the decoherence function considering the gravitons to be initially in the thermal state.}}
    \label{Fig:thermal_state}
\end{figure}

\begin{figure}[!ht]
    \centering
    \includegraphics[width=\linewidth]{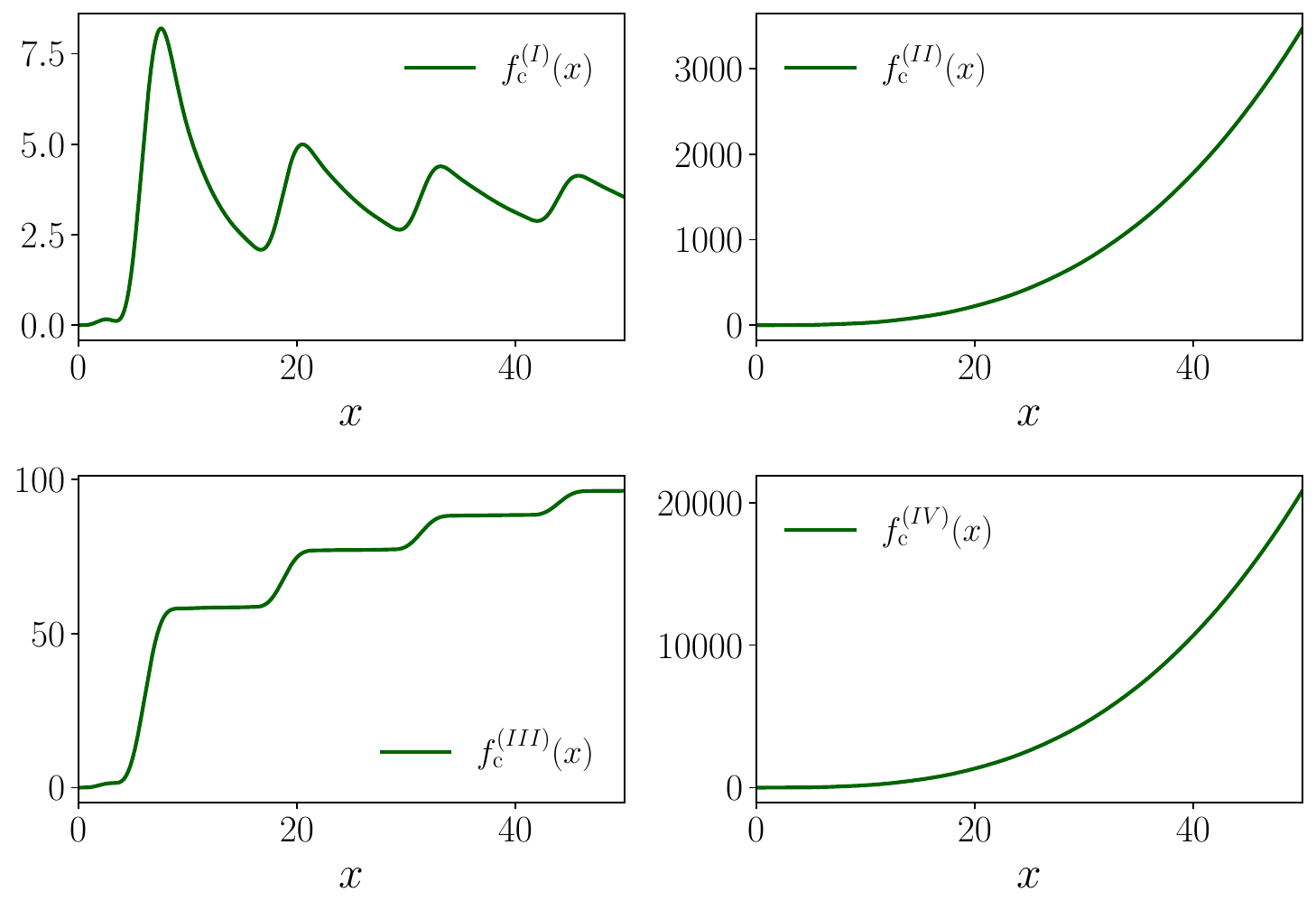}
    \caption{\justifying{Different contributions for the decoherence function considering the gravitons to be initially in the coherent state.}}
    \label{Fig:coherent_state}
\end{figure}

\begin{figure}[!ht]
    \centering
    \includegraphics[width=\linewidth]{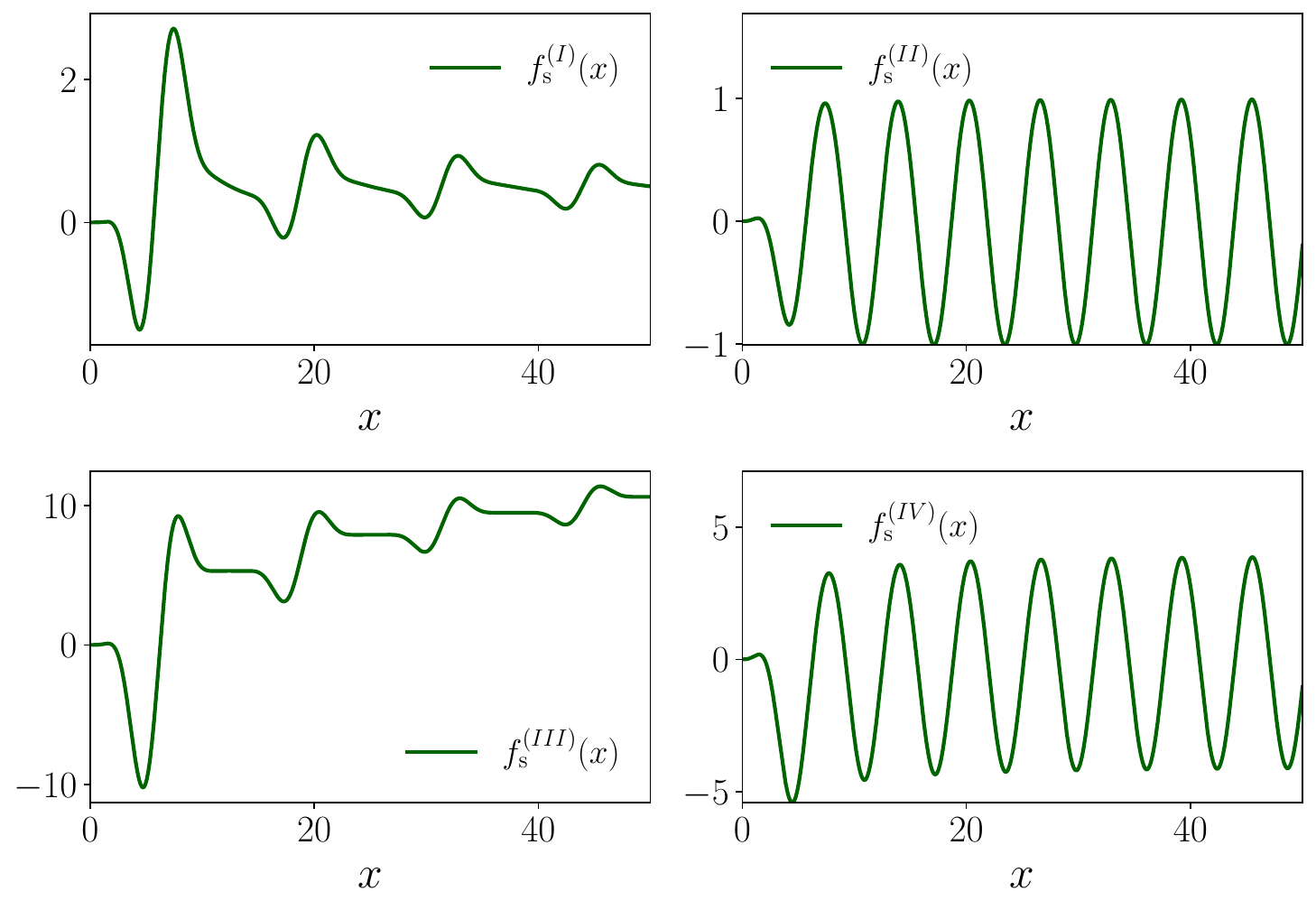}
    \caption{\justifying{Different contributions for the decoherence function considering the gravitons to be initially in the squeezed state.}}
    \label{Fig:squeezed_state}
\end{figure}

Let us next repeat the analysis for the other initial graviton states. We introduce the index $A$ that can be t, c, or s, representing thermal, coherent, and squeezed states, respectively, so that the decoherence function can be written as
\begin{align}
    &\Gamma_1^{(A)}(t)=b_A\Gamma_1^{\rm (v)}(t)+\frac{8\Xi^2v^2m^2}{5\pi E_{\rm P}^2}K_{1,A} \nonumber \\
    &\times\left\{ \qty(\frac{\Lambda_A}{\hbar})^2\qty[f_A^{(I)}\qty(\frac{\Lambda_At}{\hbar})+\frac{\kappa_A}{(mc^2)^2}f_A^{(II)}\qty(\frac{\Lambda_At}{\hbar})]\right. \nonumber \\
    &\hspace{0.5cm}\left. -\frac{GM_N}{R_N^3}\qty[f_A^{(III)}\qty(\frac{\Lambda_At}{\hbar})+\frac{\kappa_A}{(mc^2)^2}f_A^{(IV)}\qty(\frac{\Lambda_At}{\hbar})]\right\} .
\end{align}
Here,
\begin{subequations}
\begin{equation}
    b_A=\left\{ 
    \begin{array}{ll}
        1 & \textrm{for $A\neq\textrm{s}$} \\
        \cosh2r & \textrm{for $A=\textrm{s}$}
    \end{array}
    \right. ,
\end{equation}
\begin{equation}
    \Lambda_A=\left\{ 
    \begin{array}{ll}
        \Lambda & \textrm{for $A\neq\textrm{t}$} \\
        \pi k_BT_{\rm g} & \textrm{for $A=\textrm{t}$}
    \end{array}
    \right. ,
\end{equation}
\end{subequations}
and $\kappa_A=\eta\pi k_BT_{\rm int}\Lambda_A$, $K_{1,\textrm{t}}=4/3$, $K_{1,\textrm{c}}=\alpha^2/3$, $K_{1,\textrm{s}}=-(2/3)\sinh2r$, with $T_{\rm g}$ being the graviton temperature, $\alpha$ being the displacement parameter and  $r$ being the squeeze parameter. The functions $f_A$ are all listed in Appendix \ref{App:Explicit-expressions-dec-func}, and some of their relevant properties are shown in Tables \ref{tab:thermal_state}, \ref{tab:coherent_state} and \ref{tab:squeezed_state} (see also Figures~\ref{Fig:thermal_state}, \ref{Fig:coherent_state} and~\ref{Fig:squeezed_state}).

\begin{table*}[!ht]
\centering 

\begin{tabular}{llcc}
\toprule 

\multicolumn{4}{c}{\textbf{Thermal state}} \\
\cmidrule(lr){1-4} 

\textbf{Contribution} & \textbf{Function} & \textbf{Behavior for $x\ll1$} & \textbf{Behavior for $x\gg1$} \\
\midrule 

G        & $f_{\rm t}^{(I)}(x)$            & $\frac{x^4}{126}+O(x^5)$          & $1+O(1/x)$      \\
G+I      & $f_{\rm t}^{(II)}(x)$           & $\frac{4x^3}{189}$                & $\frac{4x^3}{189}$ \\
G+N      & $f_{\rm t}^{(III)}(x)$          & $\frac{x^4}{60}+O(x^5)$           & $4\ln2+4x-12\ln x+O(1/x)$ \\
G+N+I    & $f_{\rm t}^{(IV)}(x)$           & $\frac{2x^3}{45}$                 & $\frac{2x^3}{45}$ \\

\bottomrule 
\end{tabular}

\caption{\justifying{Different contributions for the decoherence function considering the gravitons to be initially in the thermal state.}}
\label{tab:thermal_state}

\end{table*}

\begin{table*}[!ht]
\centering 

\begin{tabular}{llcc}
\toprule 

\multicolumn{4}{c}{\textbf{Coherent state}} \\
\cmidrule(lr){1-4} 

\textbf{Contribution} & \textbf{Function} & \textbf{Behaviour for $x\ll1$} & \textbf{Behaviour for $x\gg1$} \\
\midrule 

G        & $f_{\rm c}^{(I)}(x)$            & $\frac{x^4}{48}+O(x^5)$          & $\frac{7}{2}+O(1/x)$      \\
G+I      & $f_{\rm c}^{(II)}(x)$           & $\frac{x^3}{18}+O(x^4)$          & $\frac{x^3}{36}+\sin x+O(1/x)$ \\
G+N      & $f_{\rm c}^{(III)}(x)$          & $\frac{x^4}{8}+O(x^5)$           & $28\gamma_E+16\ln3-68\ln2+28\ln x+O(1/x)$ \\
G+N+I    & $f_{\rm c}^{(IV)}(x)$           & $\frac{x^3}{3}+O(x^4)$           & $\frac{x^3}{6}+4\sin x+O(1/x)$ \\

\bottomrule 
\end{tabular}

\caption{\justifying{Different contributions for the decoherence function considering the gravitons to be initially in the coherent state.}}
\label{tab:coherent_state}

\end{table*}

\begin{table*}[!ht]
\centering 

\begin{tabular}{llcc}
\toprule 

\multicolumn{4}{c}{\textbf{Squeezed state}} \\
\cmidrule(lr){1-4} 

\textbf{Contribution} & \textbf{Function} & \textbf{Behavior for $x\ll1$} & \textbf{Behavior for $x\gg1$} \\
\midrule 

G        & $f_{\rm s}^{(I)}(x)$            & $\frac{x^4}{96}+O(x^5)$          & $\frac{1}{2}+O(1/x)$      \\
G+I      & $f_{\rm s}^{(II)}(x)$           & $\frac{x^3}{36}+O(x^4)$          & $\sin x+O(1/x)$ \\
G+N      & $f_{\rm s}^{(III)}(x)$          & $\frac{x^4}{16}+O(x^5)$          & $4\gamma_E+4\ln\qty(\frac{81}{2^9})+4\ln x+O(1/x)$ \\
G+N+I    & $f_{\rm s}^{(IV)}(x)$           & $\frac{x^3}{6}+O(x^4)$           & $4\sin x+O(1/x)$ \\

\bottomrule 
\end{tabular}

\caption{\justifying{Different contributions for the decoherence function considering the gravitons to be initially in the squeezed state.}}
\label{tab:squeezed_state}

\end{table*}

\emph{Short-time limit}. For $t\ll\hbar/\Lambda_A$, we can use the expansions for the functions $f_A(x)$ around $x=0$ that are shown in Tables~\ref{tab:thermal_state}, \ref{tab:coherent_state} and~\ref{tab:squeezed_state}. Once again, the purely graviton bath contribution is found to dominate in this regime, and one finds the decoherence times to be given by
\begin{subequations}
\begin{equation} \label{Short-time-dec-time-thermal}
    t_{\rm dec}^{(\textrm{t})}=\qty[1+\frac{32}{21}\delta\Omega_{\rm t}\frac{(\pi k_B T_{\rm g})^6}{\Lambda^6}]^{-1/4}t_{\rm dec}^{(\textrm{v})},
\end{equation}
\begin{equation} \label{Short-time-dec-time-coherent}
    t_{\rm dec}^{(\textrm{c})}=\qty(1+\alpha^2)^{-1/4}t_{\rm dec}^{(\textrm{v})},
\end{equation}
\begin{equation} \label{Short-time-dec-time-squeezed}
    t_{\rm dec}^{(\textrm{s})}=e^{-r/2}t_{\rm dec}^{(\textrm{v})},
\end{equation}
\end{subequations}
where
\begin{equation}
    \delta\Omega_{\rm t}\equiv1+6\qty(\qty(\frac{\hbar}{\Lambda})^2-\frac{7}{20}\qty(\frac{\hbar}{\pi k_BT_{\rm g}})^2)\frac{GM_N}{R_N^3},
\end{equation}
and with $t_{\rm dec}^{(\textrm{v})}$ given by Eq.~\eqref{Short-time-dec-time-vacuum}. We note that there is a decrease in the decoherence time when considering such states, with $t_{\rm dec}$ even exhibiting an exponential decay for the squeezed state.

\emph{Long-time limit}. Similarly to the case of the initial vacuum state, for the thermal and coherent states, as we can see from Tables~\ref{tab:thermal_state} and~\ref{tab:coherent_state}, the contributions of G+I and G+N+I continue to increase to infinity so as to eventually dominate over the others for $\Lambda_A t/\hbar\gg1$. Then, in this regime, we find the decoherence times to be given by
\begin{subequations}
\begin{equation} \label{Long-time-dec-time-thermal}
    \tau_{\rm dec}^{(\textrm{t})}=\qty[1+\frac{32}{21}\delta\Omega_{\rm t}\frac{(\pi k_B T_{\rm g})^6}{\Lambda^6}]^{-1/3}\tau_{\rm dec}^{(\textrm{v})},
\end{equation}
\begin{equation} \label{Long-time-dec-time-coherent}
    \tau_{\rm dec}^{(\textrm{c})}=\qty(1+\frac{\alpha^2}{2})^{-1/3}\tau_{\rm dec}^{(\textrm{v})},
\end{equation}
\end{subequations}
with $\tau_{\rm dec}^{(\textrm{v})}$ given by Eq.~\eqref{Long-time-dec-time-vacuum}. Once again we see that there is a decrease in decoherence time for such initial graviton states.

However, the scenario changes when considering the initial squeezed state. For long times, the contributions coming from the interplay between the gravitons and the internal DoFs behave as periodic functions which do not increase. Due to the typical smallness of the ratio $\mathscr{R}$, the contributions coming from the graviton bath alone will also dominate in the long-time limit. However, it is important to remark that we are referring only to the additional contribution to the decoherence function for the squeezed state, and one should keep in mind that for all states, this is added to the vacuum term, for which we found the G+I contributions to dominate for long times.

Now, $f_{\rm s}^{(I)}(x)$ saturates at a constant value. Furthermore, although there is a logarithmic increase in $f_{\rm s}^{(III)}(x)$, this is still much smaller than the vacuum contribution, which scales as $x^3$. Therefore, the long-time limit simply reads
\begin{equation}
    \Gamma_1^{(\textrm{s})}(t)=(\cosh2r)\Gamma_1^{(\textrm{v})}(t),
\end{equation}
with $\Gamma_1^{(\textrm{v})}(t)$ given by Eq.~\eqref{Final-dec-func-long-time}. Finally, the decoherence time is
\begin{equation} \label{Long-time-dec-time-squeezed}
    \tau_{\rm dec}^{(\textrm{s})}=\qty(\cosh2r)^{-1/3}\tau_{\rm dec}^{(\textrm{v})}.
\end{equation}
Once again, we find an exponential decay for the decoherence time with respect to the vacuum state. In Ref.~\cite{Grishchuk1990}, the authors argue that relic gravitons should now be in squeezed states, and they also estimate a squeeze parameter of order $r\sim10^2$, for which $\qty(\cosh2r)^{-1/3}\sim10^{-29}$, representing a significant reduction in the graviton-induced decoherence time.

\subsection{Gravitational recoherence} \label{Grav-rec}

Finally, let us turn our attention to the possibility of gravitational recoherence. Recall that, when we consider only the pure graviton contribution to the decoherence function, it saturates at a value given by Eq.~\eqref{Gamma_sat} (for the vacuum state). This is no longer the case when we include the Newtonian potential, since the contribution coming from the interplay between the gravitons and the classical potential does not saturate to a fixed value, but keeps increasing with time. For simplicity, let us now consider a particle with no dynamical internal DoFs, such that the decoherence function for the vacuum state becomes
\begin{align}
    \Gamma_1^{(\textrm{v})}(t)&=\frac{16\Xi^2v^2m^2}{5\pi E_{\rm P}^2} \nonumber \\
    &\times\left\{ \qty(\frac{\Lambda}{\hbar})^2f_{\rm v}^{(I)}\qty(\frac{\Lambda t}{\hbar})-\frac{GM_N}{R_N^3}f_{\rm v}^{(III)}\qty(\frac{\Lambda t}{\hbar})\right\} .
\end{align}

In the short-time limit, this expression reduces to Eq.~\eqref{Final-dec-func-short-time}, but now there is no internal DoFs contribution to dominate in the long-time regime. From Table~\ref{tab:vacuum_state}, we see that $f_{\rm v}^{(I)}(x)\sim1$ and $f_{\rm v}^{(III)}(x)\sim8\ln x$ for $x\gg1$, and the decoherence function for $t\gg\hbar/\Lambda$ becomes
\begin{align}
    \Gamma_1^{(\textrm{v})}(t)&\simeq\frac{16}{5\pi}\qty(\frac{v}{c})^2\qty(\frac{m}{M_{\rm P}})^2 \nonumber \\
    &\times\qty[1-8\qty(\frac{\hbar}{\Lambda})^2\frac{GM_N}{R_N^3}\ln\qty(\frac{\Lambda t}{\hbar})].
\end{align}
This means that, when there are no internal DoFs to couple with the gravitons and guarantee gravitational decoherence, not only the decoherence function stops increasing with time but also starts to decrease, with the possibility of eventually becoming negative (and thus no longer being called a \emph{decoherence} function). This can happen, at least in principle, and a quick analysis shows that $\Gamma_1^{(\textrm{v})}(t)<0$ for
\begin{equation}
    t\gtrsim\frac{\hbar}{\Lambda}\exp\qty[\frac{1}{8}\qty(\frac{\Lambda}{\hbar})^2\frac{R_N^3}{GM_N}].
\end{equation}
A quick numerical estimate shows that this is typically an infinitely long time. For example, even for a superposition size on the order of kilometres, $L_0\sim10^3$ m, we find $\qty(\frac{\Lambda}{\hbar})^2\frac{R_N^3}{GM_N}\sim10^{16}$ near Earth.

For the coherent and squeezed states, the situation is similar, since their G+N contributions also increase as $\ln x$ for $x\gg1$. However, for the thermal state, we see from Table~\ref{tab:thermal_state} that $f_{\rm t}^{(III)}(x)\sim4x$. Therefore, for a particle with no dynamical internal DoFs, the thermal decoherence function in the long-time limit reads
\begin{align}
    \Gamma_1^{(\textrm{t})}(t)&\simeq\frac{16}{5\pi}\qty(\frac{v}{c})^2\qty(\frac{m}{M_{\rm P}})^2 \nonumber \\
    &\times\qty[1+\qty(\frac{\pi k_B T_{\rm g}}{\Lambda})^2\qty(1-\frac{4\hbar}{\pi k_B T_{\rm g}}\frac{GM_N}{R_N^3}t)],
\end{align}
and condition $\Gamma_1^{(\textrm{t})}(t)<0$ is achieved by
\begin{equation}
    t\gtrsim\frac{\pi k_BT_{\rm g}}{4\hbar}\frac{R_E^3}{GM_E}\qty[1+\qty(\frac{\Lambda}{\pi k_B T_{\rm g}})^2].
\end{equation}
Even though the behaviour is slightly different for the thermal initial state, this result also translates to an infinitely long time. For $T_{\rm g}\sim1$ K, we have $\frac{\pi k_BT_{\rm g}}{4\hbar}\frac{R_N^3}{GM_N}\sim10^{17}$ s (near Earth), which is about the age of the universe.

Our results indicate that the presence of the Newtonian potential contributes to slightly slowing down the decoherence mechanism, which eventually can lead to recoherence on the system. In general, recoherence mechanisms can appear in non-Markovian models for open quantum systems, such as when one describes quantum fields in cosmological backgrounds.~\cite{Colas2023}. The connection between memory (non-Markovian) effects and recoherence is explored in~\cite{Kranas2025}.

The possible recoherence induced by the Newtonian contribution should be interpreted with caution. Within the perturbative treatment adopted here and in the absence of dynamical internal degrees of freedom, the Newtonian correction can formally reduce the decoherence function and eventually make it negative. However, the corresponding timescale is extremely large for realistic parameters, being comparable to cosmological timescales in the estimates considered above. Moreover, once the dynamical internal degrees of freedom are included, the long-time contribution arising from their interplay with the graviton bath dominates and decoherence eventually prevails. Thus, we do not interpret this recoherence as an experimentally accessible prediction, but rather as a formal indication of how the classical tidal background modifies the graviton-induced influence functional. The main role of the Newtonian potential in the present work is therefore to quantify the effect of a classical gravitational background on the decoherence mechanism, not to propose an observable recoherence mechanism.

Let us finally comment on the possible experimental relevance of the above results. For parameters typical of present matter-wave and optomechanical platforms, such as molecule interferometry, levitated nanoparticles and mechanical resonators, the decoherence times obtained from the present gravitational mechanism are extremely long when compared with both the coherence times currently achieved experimentally and the timescales associated with ordinary non-gravitational sources of decoherence. Thus, although the coupling to internal degrees of freedom can strongly enhance graviton-induced decoherence relative to the point-particle case, the absolute magnitude of the effect remains far too small to be observed in forthcoming tabletop experiments. A realistic observation of this effect would require parameter regimes far beyond those presently available, involving substantially larger masses, longer coherent evolution times, larger effective superposition sizes, or enhanced gravitational-noise backgrounds. The main significance of the present work is therefore conceptual. It shows that the internal dynamical structure of a composite system changes the long-time behaviour of graviton-induced decoherence and can prevent the saturation found in the point-particle case.

\section{Conclusions} \label{Sec:Conclusions}

The unavoidability of the quantum spacetime fluctuations, here understood as a graviton bath, was explored in the context of decoherence of a composite particle. We showed that while the interaction with the gravitons alone dominates for short times, the spacetime-induced coupling with the internal degrees of freedom of the system inevitably leads to gravitational decoherence for long times. This effect was also found to be enhanced by different initial graviton states, with the decoherence time even exhibiting an exponential decay for an initial squeezed state.

Although previous results in the literature indicated that graviton-induced decoherence occurs only for a macroscopic system~\cite{Kanno2021}, we showed that the situation can change when we consider composite systems. Since gravity also couples with the dynamical internal degrees of freedom, which themselves are to be treated as another environment for the centre-of-mass variable, their interaction amounts to an amplification of such a decoherence mechanism. We then raise the possibility of considering more general scenarios in which a system simultaneously interacts with both a gravitational and a non-gravitational environment, like an electromagnetic bath, for instance. Since environments will inevitably interact with each other, one may wonder what the effects of such an interplay on the system are. We leave the question of non-gravitational environments being used to mediate and amplify gravitational decoherence, just as the internal structure of the system, for future investigation.

In this work, we also considered the effects of a classical Newtonian potential, which not only contributes to slightly slowdown decoherence but also opens up the possibility of a recoherence mechanism if the internal dynamics is turned off. Such an effect usually appears in non-Markovian models for open quantum systems, and, although gravitational recoherence is shown to be possible in principle, it seems to happen for timescales comparable to the age of the universe. Nevertheless, decoherence will prevail for general composite systems. It is a matter for future investigation to determine if other classical backgrounds beyond the Newtonian approximation, such as the Schwarzschild solution or the Robertson-Walker expanding universe, will have a significant impact for the time evolution of composite open quantum systems. The dynamical nature of the metric may provide conditions that protect the system from decoherence~\cite{Pinto2011,Celeri2008}.

Finally, we remark that we considered a nonrelativistic particle in this work, and also in Ref.~\cite{Moreira_2023}, but it would be interesting to see how these extend to relativistic particles and quantum fields. For example, one could study how (amplified) graviton-induced decoherence affects neutrino oscillations through its effects on the propagation of neutrino wave packets~\cite{D_Esposito_2024,Domi2024}.

\section*{Acknowledgements} This research was funded by CNPq through grant 308065/2022-0, the National Institute of Science and Technology for Applied Quantum Computing through CNPq grant 408884/2024-0, the Coordination of Superior Level Staff Improvement (CAPES), and FAPEG through grant 202510267001843.


\onecolumngrid

\appendix

\section{Gravitational waves interacting with a Newtonian potential} \label{A:Gravitons-Newton}

In this appendix, we compute the action describing the interaction between gravitational waves (which lead to gravitons upon quantisation) and a classical Newtonian potential.

\subsection{Linearized Einstein equation}

Consider a given background metric $\Tilde{g}_{\mu\nu}$ that is an exact known solution to Einstein's equation. In situations where there is a small deviation from such exact solution, one can look for an approximate solution by writing
\begin{equation} \label{g=tildeg+h}
    g_{\mu\nu}=\Tilde{g}_{\mu\nu}+h_{\mu\nu},\hspace{0.5cm}\abs{h_{\mu\nu}}\ll\abs{\Tilde{g}_{\mu\nu}},
\end{equation}
linearising Einstein's equation in $h_{\mu\nu}$. The metric perturbation $h_{\mu\nu}$ satisfies the linear equation~\cite{Wald1984}
\begin{equation} \label{Linearized-Einstein-equation}
    \Tilde{\nabla}_{(\nu}\Tilde{\nabla}^\mu\Bar{h}_{\sigma)\mu}-\frac{1}{2}\Tilde{\nabla}^\mu\Tilde{\nabla}_\mu h_{\sigma\nu}-\Tilde{R}_{\rho\sigma\mu\nu}h^{\rho\mu}=0,
\end{equation}
where
\begin{equation}
    \Bar{h}_{\mu\nu}\equiv h_{\mu\nu}-\frac{1}{2}\Tilde{g}_{\mu\nu}h,
\end{equation}
with $h={h^\mu}_\mu=\Tilde{g}^{\mu\nu}h_{\mu\nu}$. Here, $\Tilde{\nabla}_\mu$ and $\Tilde{R}_{\rho\sigma\mu\nu}$ denote the covariant derivative and the Riemann curvature tensor associated with the background metric $\Tilde{g}_{\mu\nu}$, respectively.

Now, the gauge invariance of linearised theory states that $h_{\mu\nu}$ and $h_{\mu\nu}'$ represent the same physical perturbation, where~\cite{Wald1984}
\begin{equation} \label{Gauge-transformation}
    h_{\mu\nu}'=h_{\mu\nu}+\Tilde{\nabla}_\mu\xi_\nu+\Tilde{\nabla}_\nu\xi_\mu,
\end{equation}
for some arbitrary vector $\xi_\mu$. This gauge freedom allows us to simplify the equations of motion for $h_{\mu\nu}$. For example, starting with any given $h_{\mu\nu}$, such that $\Tilde{\nabla}^\nu\Bar{h}_{\mu\nu}\neq0$ in general, one can perform a gauge transformation according to Eq. \eqref{Gauge-transformation}. For the new perturbation, we will have $\Tilde{\nabla}^\nu\Bar{h}'_{\mu\nu}=\Tilde{\nabla}^\nu\Bar{h}_{\mu\nu}+\Tilde{\nabla}^\nu\Tilde{\nabla}_\nu\xi_\mu$. We may then choose $\xi^\mu$ to satisfy $\Tilde{\nabla}^\nu\Tilde{\nabla}_\nu\xi_\mu=-\Tilde{\nabla}^\nu\Bar{h}_{\mu\nu}$, such that, for the perturbation in the new gauge, we now have $\Tilde{\nabla}^\nu\Bar{h}_{\mu\nu}=0$, where we changed $h'\to h$ for notational convenience. In this gauge, the linearised Einstein equation \eqref{Linearized-Einstein-equation} reads
\begin{equation} \label{Gauged-linearized-Einstein-equation}
    \Tilde{\nabla}^\mu\Tilde{\nabla}_\mu h_{\sigma\nu}+2\Tilde{R}_{\rho\sigma\mu\nu}h^{\rho\mu}=0,
\end{equation}
with the metric perturbation satisfying the gauge condition,
\begin{equation} \label{Gauge-condition}
    \Tilde{\nabla}^\nu h_{\mu\nu}-\frac{1}{2}\Tilde{\nabla}_\mu h=0.
\end{equation}

In the limit of flat spacetime, this equation reduces to $\partial^\mu\partial_\mu h_{\sigma\nu}=0$. \emph{In that limit}, one can then perform a further gauge transformation parametrised by a vector $\varepsilon^\nu$ as long as $\partial^\mu\partial_\mu\varepsilon_\nu=0$, so that both the gauge condition and the equation of motion remain satisfied. Then, one uses this residual freedom to choose $h=0$ and $h_{0i}=0$, which, together with the equations of motion, implies $h_{00}=0$ for suitable boundary conditions. This is the usual \emph{transverse-traceless} (TT) gauge choice.

However, in the case of curved background, this is generally not possible. We can try to perform such an additional gauge transformation parametrized by a vector $\varepsilon^\nu$ satisfying $\Tilde{\nabla}^\mu\Tilde{\nabla}_\mu\varepsilon_\nu=0$ (which is necessary to keep Eq. \eqref{Gauge-condition} satisfied), but now the left-hand side of the equation of motion \eqref{Gauged-linearized-Einstein-equation} transforms to
\begin{equation} \label{GaugedGauged}
    \Tilde{\nabla}^\mu\Tilde{\nabla}_\mu h'_{\sigma\nu}+2\Tilde{R}_{\rho\sigma\mu\nu}h'^{\rho\mu}=\Tilde{\nabla}^\mu\Tilde{\nabla}_\mu h_{\sigma\nu}+2\Tilde{R}_{\rho\sigma\mu\nu}h^{\rho\mu}-\qty[\Tilde{\nabla}^\mu\qty(\Tilde{R}_{\mu\sigma\rho\nu}+\Tilde{R}_{\rho\sigma\mu\nu})]\varepsilon^\rho.
\end{equation}
Clearly, the last term on the right-hand side prevents us from making such gauge transformation.

Let us then see what happens with this term when the background itself can be seen as a small perturbation of Minkowski spacetime,
\begin{equation} \label{Tilde-g}
    \Tilde{g}_{\mu\nu}=\eta_{\mu\nu}+h^{(B)}_{\mu\nu}.
\end{equation}
A direct calculation yields
\begin{equation} \label{Annoying-term}
\begin{split}
    \Tilde{\nabla}^\mu\qty(\Tilde{R}_{\rho\sigma\mu\nu}+\Tilde{R}_{\mu\sigma\rho\nu})&=\frac{1}{2}\partial_\sigma\Box h^{(B)}_{\rho\nu}+\frac{1}{2}\partial_\nu\partial_\rho\partial^\mu h^{(B)}_{\mu\sigma}+\frac{1}{2}\partial_\rho\partial_\sigma\partial^\mu h^{(B)}_{\mu\nu}+\frac{1}{2}\partial_\nu\Box h^{(B)}_{\rho\sigma} \\
    &-\partial_\rho\Box h^{(B)}_{\nu\sigma}-\partial_\nu\partial_\sigma\partial^\mu h^{(B)}_{\rho\mu}.
\end{split}
\end{equation}
Since the background metric is assumed to satisfy the vacuum equation, $h^{(B)}_{\mu\nu}$ must satisfy the linearized equation \cite{Carroll}
\begin{equation} \label{Linearized-Mink-vacuum-equation}
    \Box h^{(B)}_{\mu\nu}=\partial^\sigma\partial_\nu h^{(B)}_{\sigma\mu}+\partial^\sigma\partial_\mu h^{(B)}_{\sigma\nu}-\partial_\mu\partial_\nu h^{(B)}.
\end{equation}
By plugging Eq. \eqref{Linearized-Mink-vacuum-equation} into Eq. \eqref{Annoying-term}, one immediately finds
\begin{equation}
    \Tilde{\nabla}^\mu\qty(\Tilde{R}_{\rho\sigma\mu\nu}+\Tilde{R}_{\mu\sigma\rho\nu})=0.
\end{equation}
This means that, \emph{when the background metric can be treated as a small perturbation of Minkowski spacetime}, the term keeping us from performing the residual gauge transformation on $h_{\mu\nu}$ vanishes up to the first order in $h^{(B)}_{\mu\nu}$. Then, one is allowed to choose $\varepsilon^\mu$ satisfying $\Tilde{\nabla}^\mu\Tilde{\nabla}_\mu\varepsilon_\nu=0$, such that $h=h_{0i}=0$. The gauge condition then becomes
\begin{equation}
    \Tilde{\nabla}_\mu h^{\mu\nu}=0.
\end{equation}
In particular, the $\nu=0$ component of this equation reads
\begin{equation}
    \partial_0h^{00}+\qty(\Tilde{\Gamma}_{\mu0}^\mu+\Tilde{\Gamma}_{00}^0)h^{00}+\Tilde{\Gamma}_{ij}^0h^{ij}=0.
\end{equation}
\emph{If the metric perturbation is static}, we get $\Tilde{\Gamma}_{\mu0}^\mu=\Tilde{\Gamma}_{00}^0=0$. Additionally, \emph{if $h^{(B)}_{0i}=0$} we also have $\Tilde{\Gamma}_{ij}^0=0$, and thus
\begin{equation}
    \partial_0h^{00}=0.
\end{equation}
Then, since $h_{00}$ is just a constant, one can choose suitable boundary conditions such that $h_{00}=0$.

In summary, \emph{for a background metric that is a small static perturbation of flat spacetime with $h^{(B)}_{0i}=0$}, the Einstein equation for $h_{\mu\nu}$ reads as follows
\begin{equation} \label{Linearized-Einstein-equation-in-TT-gauge}
    \Tilde{\nabla}^\mu\Tilde{\nabla}_\mu h_{ij}+2\Tilde{R}_{ikjl}h^{kl}=0,
\end{equation}
with the TT gauge conditions
\begin{equation} \label{TT-gauge}
\begin{split}
    \Tilde{\nabla}_\mu h^{\mu\nu}=0, \\
    \Tilde{g}^{\mu\nu}h_{\mu\nu}=0, \\
    h_{0\mu}=0.
\end{split}
\end{equation}

The linearized vacuum equation in the TT gage can be obtained from the extremization of the action \cite{Chawla2023}
\begin{equation}
    S_{\rm grav}=\frac{1}{32\pi}\int\dd^4x\,\qty(\frac{1}{2}h_{ij}\Tilde{\nabla}^\mu\Tilde{\nabla}_\mu h^{ij}+h_{ij}\Tilde{R}^{ikjl}h_{kl}).
\end{equation}
We can check that setting $\delta S=0$ leads to Eq. \eqref{Linearized-Einstein-equation-in-TT-gauge}.


\subsection{Newtonian limit}

Let us see how the linearized Einstein equation works when the background is described in the Newtonian limit. In that case, we consider the expansion \eqref{Tilde-g} with $h^{(B)}_{\mu\nu}=-2\phi\delta_{\mu\nu}$, with $\phi$ being the time-independent Newtonian potential. The gravitational action is then reduced to
\begin{align} \label{Graviton-action-position-space}
    S_{\rm grav}&=\frac{1}{64\pi}\int\dd^4x\,\Tilde{g}^{\mu\nu}h_{ij}\partial_\mu\partial_\nu h^{ij} \nonumber \\
    &=\frac{1}{64\pi}\int\dd^4x\,\qty(h_{ij}\Box h^{ij}+2\phi\,h_{ij}\delta_{\mu\nu}\partial^\mu\partial^\nu h^{ij}).
\end{align}
The equation of motion for the metric perturbation, $\delta S_{\rm grav}=0$, reads
\begin{equation}
    \Box h_{ij}+2\phi\delta_{\mu\nu}\partial^\mu\partial^\nu h_{ij}=0.
\end{equation}

For a potential of the form $\phi(r)=-GM/r$, and upon quantization of the degrees of freedom of gravitational radiation, one finds that this coupling leads the mass $M$ to scatter an incident graviton by an angle $\theta$. The differential cross section can be shown to be \cite{Ragusa_2003}
\begin{equation}
    \dv{\sigma}{\Omega}=\frac{G^2M^2}{c^4\sin^4(\theta/4)}\qty(\cos^8\theta+\sin^8\theta),
\end{equation}
where we restored the universal constants $G$ and $c$.

\section{Gravitational noise kernel} \label{A:Gravitational-Noise-Kernel}

In this appendix, we compute the noise kernel encoding the influence of the quantized gravitational field, defined by Eq. \eqref{Grav-noise-kernel}, for different initial states. In what follows, we shall consider the position operators to be independent of the polarizations and of the direction of $\vb{k}$, that is, $q_s(t,\vb{k})=q_\omega(t)$, $\omega=\abs{\vb{k}}$. Within this assumption, we may write the gravitational noise kernel as
\begin{equation}
    N_{\rm g}^{ijkl}(t,t')=\frac{1}{32}\int_0^\infty\dd\omega\,\omega^2\qty[A_{\Omega}^{ijkl}\dv[2]{t}\dv[2]{{t'}}G_\omega(t,t')+A_{\Omega}^{ijnl}{T_n}^k\dv[2]{t}G_\omega(t,t')+A_{\Omega}^{njkl}{T_n}^i\dv[2]{{t'}}G_\omega(t,t')],
\end{equation}
where we introduced the Hadamard function
\begin{equation} \label{Hadamard-function}
    G_\omega(t,t')=\expval{\acomm{q_\omega(t)}{q_\omega(t')}}_{\rm g}.
\end{equation}
The angular integrals are given by \cite{Cho2022,Moreira_2023}
\begin{equation} \label{Angular-integrals}
    A_{\Omega}^{ijkl}\equiv\int \dd \Omega\sum_s\epsilon_s^{ij}(\vb{k})\epsilon_s^{kl}(\vb{k})=\frac{8\pi}{15}\qty[3\qty(\delta^{ik}\delta^{jl}+\delta^{il}\delta^{jk})-2\delta^{ij}\delta^{kl}].
\end{equation}

As we discussed in Sec. \ref{Sec:Decoherence-function}, we shall be interested only in the components $z$, $N_{\rm g}(t,t')\equiv N_{\rm g}^{3333}(t,t')$, which are then given by
\begin{equation} \label{Grav-noise-kernel-2}
    N_{\rm g}(t,t')=\frac{\pi}{15}\int_0^\infty\dd\omega\,\omega^2\qty{\dv[2]{t}\dv[2]{{t'}}G_\omega(t,t')+T_{zz}\qty[\dv[2]{t}G_\omega(t,t')+\dv[2]{{t'}}G_\omega(t,t')]}.
\end{equation}

Next, let us explicitly evaluate the Green's function \eqref{Hadamard-function} for different initial states. Since the free Hamiltonian for each mode of the gravitational field is the one of a harmonic oscillator, Eq. \eqref{Hamiltonian-individual-mode}, we may write the position operators in the Heisenberg picture as
\begin{equation}
    q_\omega(t)=\sqrt{\frac{1}{2m_{\rm g}\omega}}\qty(a_\omega e^{-i\omega t}+a_\omega^\dagger e^{i\omega t}),
\end{equation}
where the $a$'s ($a^\dagger$'s) are annihilation (creation) operators satisfying the usual commutation relations,
\begin{equation}
\begin{split}
    \comm{a_\omega}{a_{\omega'}}&=\comm{a_\omega^\dagger}{a_{\omega'}^\dagger}=0, \\
    \comm{a_\omega}{a_{\omega'}^\dagger}&=\delta_{\omega\omega'}.
\end{split}
\end{equation}
Then a direct calculation yields
\begin{equation} \label{G(t,t')}
    G_\omega(t,t')=\frac{2}{m_{\rm g}\omega^2}\expval{H_\omega}_{\rm g}\cos\omega(t-t')+\frac{1}{m_{\rm g}\omega}\qty[\expval{a_\omega^2}_{\rm g}e^{-i\omega(t+t')}+\expval{\qty(a^\dagger_\omega)^2}_{\rm g}e^{i\omega(t+t')}],
\end{equation}
where
\begin{equation}
    H_\omega=\omega\qty(a_\omega^\dagger a_\omega+\frac{1}{2})=\frac{\omega}{2}\acomm{a_\omega}{a_\omega^\dagger}
\end{equation}
is the free Hamiltonian operator of the harmonic oscillator with frequency $\omega$. We may now compute the noise kernel for different initial states of the gravitational field. These were explicitly computed in~\cite{Moreira_2023}, so here we shall only list the results.

\begin{subequations} \label{Green-function-different-states}
If the initial state is the vacuum state, one finds
\begin{equation} \label{Green-function-vacuum}
    G_\omega^{\rm (vac)}(t,t')=\frac{1}{m_{\rm g}\omega}\cos{\omega(t-t')}.
\end{equation}
Alternatively, if we take the gravitons to be initially in a thermal state with temperature $T_{\textrm{g}}$, we have
\begin{equation} \label{Green-function-thermal}
    G_\omega^{\rm (th)}(t,t')=G_\omega^{\rm (vac)}(t,t')+\frac{2}{m_{\rm g}\omega}\frac{1}{e^{\omega/T_{\rm g}}-1}\cos{\omega(t-t')}.
\end{equation}
For gravitons initially in a coherent state with displacement parameter $\alpha$, the Green's function reads
\begin{equation} \label{Green-function-coherent}
    G_\omega^{\rm (coh)}(t,t')=G_\omega^{\rm (vac)}(t,t')+\frac{\alpha^2}{m_{\rm g}\omega}\cos(\omega t)\cos(\omega t').
\end{equation}
Finally, for gravitons initially in a squeezed state with squeeze parameter $\zeta=re^{i\varphi}$, we obtain
\begin{equation} \label{Green-function-squeezed}
    G_\omega^{\rm (sq)}(t,t')=(\cosh{2r})G_\omega^{\rm (vac)}(t,t')-\frac{\sinh{2r}}{m_{\rm g}\omega}\cos[\omega(t+t')-\varphi].
\end{equation}
\end{subequations}

Our next task is then to simply plug in the results of Eqs. \eqref{Green-function-different-states} into Eq. \eqref{Grav-noise-kernel-2} for the gravitational noise kernel. For instance, we have, for the initial vacuum state,
\begin{align} \label{N-vac(t,t')}
    N_{\rm g}^{\rm (vac)}(t,t')&=\frac{2}{15\pi}\qty[\int_0^\Lambda\dd\omega\,\omega^5\cos{\omega(t-t')}-2T_{zz}\int_0^\Lambda\dd\omega\,\omega^3\cos{\omega(t-t')}] \nonumber \\
    &=\frac{2\Lambda^4}{15\pi}\qty{\Lambda^2F_5\qty[\Lambda(t-t')]-2T_{zz}F_3\qty[\Lambda(t-t')]},
\end{align}
where we introduced the frequency cutoff $\Lambda$ and used $m_{\rm g}=\pi^2/2$. Also, we define
\begin{equation} \label{Fn(x)}
    F_n(x)\equiv\frac{1}{x^{n+1}}\int_0^x\dd y\,y^n\cos y.
\end{equation}
Explicitly,
\begin{subequations} \label{F5(x)eF3(x)}
\begin{equation} \label{F5(x)}
    F_5(x)=\frac{1}{x^6}\left[ \qty(5x^4-60x^2+120)\cos x+x\qty(x^4-20x^2+120)\sin x-120\right] ,
\end{equation}
and
\begin{equation} \label{F3(x)}
    F_3(x)=\frac{1}{x^4}\qty[\qty(3x^2-6)\cos x+\qty(x^3-6x)\sin x+6].
\end{equation}
\end{subequations}

Additionally, it is easy to show that
\begin{equation} \label{Limit-Fn(x)-x-to-0}
    \lim_{x\to0}F_n(x)=\frac{1}{n+1},
\end{equation}
and thus
\begin{equation} \label{N-vac(t)}
    N^{\rm (vac)}_{\rm g}(t)=\lim_{t'\to t}N^{\rm (vac)}_{\rm g}(t,t')=\frac{\Lambda^4}{15\pi}\qty(\frac{\Lambda^2}{3}-T_{zz}).
\end{equation}

Now, for the thermal state, we find
\begin{align} \label{N-th(t,t')}
    N^{\rm (th)}_{\rm g}(t,t')&=N^{\rm (vac)}_{\rm g}(t,t')+\frac{4}{15\pi}\int_0^\infty\dd\omega\qty[\frac{\omega^5}{e^{\omega/T_{\rm g}}-1}\cos\omega(t-t')-2T_{zz}\frac{\omega^3}{e^{\omega/T_{\rm g}}-1}\cos\omega(t-t')] \nonumber \\
    &=N^{\rm (vac)}_{\rm g}(t,t')+\frac{8\pi^3T_{\rm g}^4}{5}\qty{10\pi^2T_{\rm g}^2F_1^{(\rm th)}\qty[\pi T_{\rm g}(t-t')]-T_{zz}F_2^{(\rm th)}\qty[\pi T_{\rm g}(t-t')]},
\end{align}
with
\begin{subequations} \label{Fth1-e-2}
\begin{equation} \label{Fth1}
    F_1^{(\rm th)}(x)=\frac{1}{x^6}-\frac{1}{15\sinh^6x}\qty(2\cosh^4x+11\cosh^2x+2),
\end{equation}
and
\begin{equation} \label{Fth2}
    F_2^{(\rm th)}(x)=\frac{1}{3\sinh^4x}\qty(2\cosh^2x+1)-\frac{1}{x^4}.
\end{equation}
\end{subequations}
Also, using $\lim_{x\to0}F_1^{(\rm th)}(x)=2/945$ and $\lim_{x\to0}F_2^{(\rm th)}(x)=1/45$, we find
\begin{equation} \label{N-th(t)}
    N^{\rm (th)}_{\rm g}(t)=\lim_{t'\to t}N^{\rm (th)}_{\rm g}(t,t')=N^{\rm (vac)}_{\rm g}(t)+\frac{8\pi^3T_{\rm g}^4}{45}\qty(\frac{4}{21}\pi^2T_{\rm g}^2-\frac{1}{5}T_{zz}).
\end{equation}

For the coherent state,
\begin{align} \label{N-coh(t,t')}
    N^{\rm (coh)}_{\rm g}(t,t')&=N^{\rm (vac)}_{\rm g}(t,t')+\frac{2\alpha^2}{15\pi}\qty[\int_0^\Lambda\dd\omega^5\cos(\omega t)\cos(\omega t')-2T_{zz}\int_0^\Lambda\dd\omega^3\cos(\omega t)\cos(\omega t')] \nonumber \\
    &=N^{\rm (vac)}_{\rm g}(t,t')+\frac{\alpha^2\Lambda^4}{15\pi}\left\{ \Lambda^2\qty[F_5\qty(\Lambda(t+t'))+F_5\qty(\Lambda(t-t'))]\right. \nonumber \\
    &\hspace{3.5cm}\left. -2T_{zz}\qty[F_3\qty(\Lambda(t+t'))+F_3\qty(\Lambda(t-t'))]\right\} ,
\end{align}
and
\begin{equation} \label{N-coh(t)}
    N^{\rm (coh)}_{\rm g}(t)=\lim_{t'\to t}N^{\rm (coh)}_{\rm g}(t,t')=N^{\rm (vac)}_{\rm g}(t)+\frac{\alpha^2\Lambda^4}{15\pi}\left\{ \Lambda^2\qty[F_5\qty(2\Lambda t)+\frac{1}{6}]-2T_{zz}\qty[F_3\qty(2\Lambda t)+\frac{1}{4}]\right\} 
\end{equation}

Lastly, for the squeezed state, assuming the squeeze parameter to be real for simplicity ($\varphi=0$), the noise kernel becomes
\begin{align} \label{N-sq(t,t')}
    N^{\rm (sq)}_{\rm g}(t,t')&=(\cosh2r)N^{\rm (vac)}_{\rm g}(t,t')-\frac{2}{15\pi}\sinh2r\qty[\int_0^\Lambda\omega^5\cos\omega(t+t')-2T_{zz}\int_0^\Lambda\omega^3\cos\omega(t+t')] \nonumber \\
    &=(\cosh2r)N^{\rm (vac)}_{\rm g}(t,t')-\frac{2\Lambda^4}{15\pi}\sinh2r\qty{\Lambda^2F_5\qty[\Lambda(t+t')]-2T_{zz}F_3\qty[\Lambda(t+t')]},
\end{align}
and
\begin{equation} \label{N-sq(t)}
    N^{\rm (sq)}_{\rm g}(t)=\lim_{t'\to t}N^{\rm (sq)}_{\rm g}(t,t')=(\cosh2r)N^{\rm (vac)}_{\rm g}(t)-\frac{2\Lambda^4}{15\pi}\sinh2r\qty[\Lambda^2F_5\qty(2\Lambda t)-2T_{zz}F_3\qty(2\Lambda t)].
\end{equation}

\section{Explicit expressions for the decoherence function} \label{App:Explicit-expressions-dec-func}

In this appendix, we list the results for the decoherence function by considering the two superposition configurations mentioned in Section \ref{Sec:Decoherence-function}, and the gravitons initially in the four different states described in Appendix~\ref{A:Gravitational-Noise-Kernel}.

\subsection{Configuration 1}

Configuration 1 is the one described by
\begin{subequations}
\begin{equation}
    \Xi(t)=\Xi=\textrm{constant in time}
\end{equation}
and
\begin{equation}
    \Delta\xi(t')=\left\{
    \begin{array}{ll}
        2vt' &\textrm{for}\hspace{0.2cm}0<t'\leq t/2\\
        2v(t-t') &\textrm{for}\hspace{0.2cm}t/2<t'<t
    \end{array}
    \right.,
\end{equation}
\end{subequations}
for some constant velocity $v$, and $t$ denotes the time span of the superposition state. The decoherence function reads
\begin{align}
    \Gamma_1(t)&=8m^2\Xi^2v^2\left[ \int_0^{t/2}\dd t_1\dd t_2\,t_1t_2N_{\rm g}(t_1,t_2)+\int_{t/2}^{t}\dd t_1\dd t_2\,(t-t_1)(t-t_2)N_{\rm g}(t_1,t_2) \right. \nonumber \\
    &\hspace{2cm}\left. +2\int_0^{t/2}\dd t_1\int_{t/2}^{t}\dd t_2\,t_1(t-t_2)N_{\rm g}(t_1,t_2)\right] \nonumber \\
    &+16\eta\pi T_{\rm int}\Xi^2v^2\qty[\int_0^{t/2}\dd t_1\,t_1^2N_{\rm g}(t_1)+\int_{t/2}^{t}\dd t_1\,(t-t_1)^2N_{\rm g}(t_1)].
\end{align}
Now we can plug in the expressions for the noise kernel (which were computed in Appendix \ref{A:Gravitational-Noise-Kernel}), use $T_{zz}=2M_N/R_N^3$ and restore the universal constants.

In order to present it in a unifying notation, let us introduce the index $A$ that can be either v, t, c or s, representing vacuum, thermal, coherent and squeezed states, respectively. Then, the results for the function $\Gamma_1(t)$ can be summarized in the following expression:
\begin{align}
    \Gamma_1^{(A)}(t)&=b_A(1-\delta_{A,\textrm{v}})\Gamma_1^{\rm (v)}(t) \nonumber \\
    &+\frac{8\Xi^2v^2m^2}{5\pi E_{\rm P}^2}K_{1,A}\left\{ \qty(\frac{\Lambda_A}{\hbar})^2\qty[f_A^{(I)}\qty(\frac{\Lambda_At}{\hbar})+\frac{\kappa_A}{(mc^2)^2}f_A^{(II)}\qty(\frac{\Lambda_At}{\hbar})]\right. \nonumber \\
    &\hspace{2.5cm}\left. -\frac{GM_N}{R_N^3}\qty[f_A^{(III)}\qty(\frac{\Lambda_At}{\hbar})+\frac{\kappa_A}{(mc^2)^2}f_A^{(IV)}\qty(\frac{\Lambda_At}{\hbar})]\right\} .
\end{align}
Here,
\begin{subequations} \label{Definitions-contants-Gamma(t)}
\begin{equation}
    b_A=\left\{ 
    \begin{array}{ll}
        1 & \textrm{for $A\neq\textrm{s}$} \\
        \cosh2r & \textrm{for $A=\textrm{s}$}
    \end{array}
    \right. ,
\end{equation}
\begin{equation}
    \Lambda_A=\left\{ 
    \begin{array}{ll}
        \Lambda & \textrm{for $A\neq\textrm{t}$} \\
        \pi k_BT_{\rm g} & \textrm{for $A=\textrm{t}$}
    \end{array}
    \right. ,
\end{equation}
and
\begin{equation}
    \kappa_A=\eta\pi k_BT_{\rm int}\Lambda_A.
\end{equation}
\end{subequations}
Also, $E_{\rm P}=\sqrt{\hbar c^5/G}\simeq2.0\times10^9\,\textrm{J}$ is the Planck energy. The constants $K_{1,A}$ are
\begin{subequations}
\begin{align}
    &K_{1,\textrm{v}}=2, \\
    &K_{1,\textrm{t}}=\frac{4}{3}, \\
    &K_{1,\textrm{c}}=\frac{\alpha^2}{3}, \\
    &K_{1,\textrm{s}}=-\frac{2}{3}\sinh2r.
\end{align}
\end{subequations}
The various functions $f_A$ are listed below:

\paragraph{Vacuum state}

\begin{subequations}
\begin{equation}
    f_{\rm v}^{(I)}(x)=1+\frac{2}{3x}\qty[\sin x-8\sin\qty(\frac{x}{2})]+\frac{1}{x^2}\qty[\frac{2}{3}\cos x-\frac{32}{3}\cos\qty(\frac{x}{2})+10],
\end{equation}
\begin{equation}
    f_{\rm v}^{(II)}(x)=\frac{1}{108}x^3,
\end{equation}
\begin{equation}
    f_{\rm v}^{(III)}(x)=8\gamma_E-\frac{4}{3}\ln4-\frac{32}{3}\textrm{Ci}\qty(\frac{x}{2})+\frac{8}{3}\textrm{Ci}(x)+8\ln\qty(\frac{x}{2}),
\end{equation}
\begin{equation}
    f_{\rm v}^{(IV)}(x)=\frac{1}{18}x^3.
\end{equation}
\end{subequations}

\paragraph{Thermal state}

\begin{subequations}
\begin{equation}
    f_{\rm t}^{(I)}(x)=\frac{1+16e^x+26e^{2x}+16e^{3x}+e^{4x}}{(e^{2x}-1)^2}-\frac{15}{x^2},
\end{equation}
\begin{equation}
    f_{\rm t}^{(II)}(x)=\frac{4}{189}x^3,
\end{equation}
\begin{equation}
    f_{\rm t}^{(III)}(x)=4\ln\qty[\frac{2(e^x-1)^3}{x^3(e^x+1)}]-4x,
\end{equation}
\begin{equation}
    f_{\rm t}^{(IV)}(x)=\frac{2}{45}x^3.
\end{equation}
\end{subequations}

\paragraph{Coherent state}

\begin{subequations}
\begin{align}
    f_{\rm c}^{(I)}(x)&=\frac{7}{2}+\frac{1}{6x}\qty{3\sin(2x)-16\qty[9\sin\qty(\frac{x}{2})-3\sin x+\sin\qty(\frac{3x}{2})]} \nonumber \\
    &+\frac{1}{36x^2}\qty[1495-1728\cos\qty(\frac{x}{2})+288\cos x-64\cos\qty(\frac{3x}{2})+9\cos(2x)],
\end{align}
\begin{align}
    f_{\rm c}^{(II)}(x)&=\frac{x^3}{36}+\frac{1}{8x^3}\left[ 49+24(x^2-2)\cos x+(2x^2-1)\cos(2x)\right. \nonumber \\
    &\left. -4x(12-2x^2+\cos x)\sin x\right] ,
\end{align}
\begin{equation}
    f_{\rm c}^{(III)}(x)=28\gamma_E+4\ln\qty(\frac{81x^7}{2^{17}})-48\textrm{Ci}\qty(\frac{x}{2})+32\textrm{Ci}(x)-16\textrm{Ci}\qty(\frac{3x}{2})+4\textrm{Ci}(2x),
\end{equation}
\begin{equation}
    f_{\rm c}^{(IV)}(x)=\frac{x^3}{6}+4\sin x+\frac{2}{x}(2\cos x+\cos^2x-3).
\end{equation}
\end{subequations}

\paragraph{Squeezed state}

\begin{subequations}
\begin{align}
    f_{\rm s}^{(I)}(x)&=\frac{1}{2}+\frac{1}{2x}\qty[\sin(2x)+12\sin x-16\sin\qty(\frac{x}{2})-\frac{16}{3}\sin\qty(\frac{3x}{2})] \nonumber \\
    &+\frac{1}{36x^2}\qty[415-576\cos\qty(\frac{x}{2})+216\cos x-64\cos\qty(\frac{3x}{2})+9\cos(2x)],
\end{align}
\begin{align}
    f_{\rm s}^{(II)}(x)&=\frac{1}{8x^3}\left[ 49+24(x^2-2)\cos x+(2x^2-1)\cos(2x)\right. \nonumber \\
    &\left. -4x(12-2x^2+\cos x)\sin x\right] ,
\end{align}
\begin{equation}
    f_{\rm s}^{(III)}(x)=4\gamma_E+4\ln\qty(\frac{81x}{2^{9}})-16\textrm{Ci}\qty(\frac{x}{2})+24\textrm{Ci}(x)-16\textrm{Ci}\qty(\frac{3x}{2})+4\textrm{Ci}(2x),
\end{equation}
\begin{equation}
    f_{\rm s}^{(IV)}(x)=4\sin x+\frac{2}{x}(2\cos x+\cos^2x-3).
\end{equation}
\end{subequations}
Here $\gamma_E\simeq0.577$ is the Euler-Mascheroni constant and $\textrm{Ci}(z)=-\int_z^\infty\frac{\cos t}{t}\dd t$ is the cosine integral function.

\subsection{Configuration 2}

Configuration 2 is described by
\begin{subequations}
\begin{equation}
    \Xi(t)=Vt,\hspace{0.5cm}\Delta\xi(t)=\Delta vt,
\end{equation}
and
\begin{equation}
    V=\frac{v_1+v_2}{2},\hspace{0.5cm}\Delta v=v_1-v_2,
\end{equation}
\end{subequations}
for which the decoherence rate becomes
\begin{equation}
    \Gamma_2(t)=\frac{1}{4}\eta\pi T_{\rm int}\qty(v_1^2-v_2^2)^2\qty[\frac{t}{2}+\frac{T_{zz}t^3}{3}+4\int_0^{t}\dd t_1\,t_1^4N_{\rm g}(t_1)]+\frac{m^2}{2}\qty(v_1^2-v_2^2)^2\int_0^{t}\dd t_1\dd t_2\,(t_1t_2)^2N_{\rm g}(t_1,t_2).
\end{equation}

Proceeding as we did for the last configuration leads to
\begin{subequations}
\begin{equation}
    \Gamma_2^{(A)}(t)=\frac{\pi\eta k_BT_{\rm int}}{4\hbar}\frac{\qty(v_1^2-v_2^2)^2}{c^4}\qty(t+\frac{2}{3}\frac{GM_E}{R_E^3}t^3)+\Gamma_{2(g)}^{(A)}(t),
\end{equation}
with
\begin{equation}
\begin{split}
    \Gamma_{2(g)}^{(A)}(t)&=b_A(1-\delta_{A,\textrm{v}})\Gamma_{2(g)}^{\rm (v)}(t) \\
    &+\frac{m^2\qty(v_1^2-v_2^2)^2}{15\pi E_{\rm P}^2}K_{2,A}\left\{ g_A^{(I)}\qty(\frac{\Lambda_At}{\hbar})+\frac{\kappa_A}{(mc^2)^2}g_A^{(II)}\qty(\frac{\Lambda_At}{\hbar})\right. \\
    &\hspace{2.5cm}\left. -\qty(\frac{\hbar}{\Lambda_A})^2\frac{GM_N}{R_N^3}\qty[g_A^{(III)}\qty(\frac{\Lambda_At}{\hbar})+\frac{\kappa_A}{(mc^2)^2}g_A^{(IV)}\qty(\frac{\Lambda_At}{\hbar})]\right\} ,
\end{split}
\end{equation}
\end{subequations}
with $b_A$, $\Lambda_A$ and $\kappa_A$ defined in Eqs. \eqref{Definitions-contants-Gamma(t)}.
The constants $K_{2,A}$ are
\begin{subequations}
\begin{align}
    &K_{2,\textrm{v}}=1, \\
    &K_{2,\textrm{t}}=12, \\
    &K_{2,\textrm{c}}=\alpha^2, \\
    &K_{2,\textrm{s}}=-\sinh2r.
\end{align}
\end{subequations}
The various functions $g_A$ are listed below:

\paragraph{Vacuum state}

\begin{subequations}
\begin{equation}
    g_{\rm v}^{(I)}(x)=\frac{x^4}{4}+8\gamma_E-12-8\textrm{Ci}(x)+8\ln x+4x\sin x+12\cos x,
\end{equation}
\begin{equation}
    g_{\rm v}^{(II)}(x)=\frac{x^5}{15},
\end{equation}
\begin{equation}
    g_{\rm v}^{(III)}(x)=2x^4-8x^2+16\cos x+16x\sin x-16,
\end{equation}
\begin{equation}
    g_{\rm v}^{(IV)}(x)=\frac{2}{5}x^5.
\end{equation}
\end{subequations}

\paragraph{Thermal state}

\begin{subequations}
\begin{equation}
    g_{\rm t}^{(I)}(x)=1-\frac{2x}{3}+\frac{x^4}{90}+\frac{2}{3}\ln\qty(\frac{e^{2x}-1}{2x})-\frac{x}{3}\qty[\frac{\sinh(2x)+x}{\sinh^2x}],
\end{equation}
\begin{equation}
    g_{\rm t}^{(II)}(x)=\frac{8}{945}x^5,
\end{equation}
\begin{equation}
    g_{\rm t}^{(III)}(x)=\frac{x^4}{9}+\frac{4}{9}x^3+\frac{2}{3}x^2-\frac{4}{3}x^2\ln(1-e^{2x})-\frac{4}{3}x\textrm{Li}_2(e^{2x})+\frac{2}{3}\textrm{Li}_3(e^{2x})-\frac{2}{3}\zeta(3),
\end{equation}
\begin{equation}
    g_{\rm t}^{(IV)}(x)=\frac{4}{225}x^5.
\end{equation}
\end{subequations}

\paragraph{Coherent state}

\begin{subequations}
\begin{equation}
    g_{\rm c}^{(I)}(x)=\frac{x^4}{8}+2\gamma_E-\frac{59}{16}+\frac{1}{16}(59-22x^2)\cos(2x)-2\textrm{Ci}(2x)-2\ln\qty(\frac{x}{2})+4\ln x+\frac{x}{8}(27-2x^2)\sin(2x),
\end{equation}
\begin{equation}
    g_{\rm c}^{(II)}(x)=\frac{x^5}{30}+\frac{1}{8x}\qty[15+\qty(-15+18x^2-2x^4)\cos(2x)]+(x^2-3)\sin(2x),
\end{equation}
\begin{equation}
    g_{\rm c}^{(III)}(x)=x^4-\frac{7}{2}x^2-4+\qty(4-\frac{9}{2}x^2)\cos(2x)-x(x^2-8)\sin(2x),
\end{equation}
\begin{equation}
    g_{\rm c}^{(IV)}(x)=\frac{3}{2}x+\frac{x^5}{5}+x\qty(\frac{9}{2}-x^2)\cos(2x)+3(x^2-1)\sin(2x).
\end{equation}
\end{subequations}

\paragraph{Squeezed state}

\begin{subequations}
\begin{equation}
    g_{\rm s}^{(I)}(x)=\frac{37}{8}-4\gamma_E-12\cos x+\frac{1}{8}(59-22x^2)\cos(2x)+8\textrm{Ci}(x)-4\textrm{Ci}(2x)-4\ln\qty(\frac{x}{2})-\frac{x}{2}\qty[8+(2x^2-27)\cos x]\sin x,
\end{equation}
\begin{equation}
    g_{\rm s}^{(II)}(x)=\frac{1}{4x}\qty[15+\qty(-15+18x^2-2x^4)\cos(2x)]+2(x^2-3)\sin(2x),
\end{equation}
\begin{equation}
    g_{\rm s}^{(III)}(x)=x^2+8-16\cos x+(8-9x^2)\cos(2x)-4x\qty[4+(x^2-8)\cos x]\sin x,
\end{equation}
\begin{equation}
    g_{\rm s}^{(IV)}(x)=3x+x(9-2x^2)\cos(2x)+6(x^2-1)\sin(2x).
\end{equation}
\end{subequations}
Here $\zeta(z)$ is the Riemann zeta function and $\textrm{Li}_n(z)$ is the polylogarithm function.



\twocolumngrid

\bibliographystyle{apsrev4-2}

\bibliography{ref.bib}

\end{document}